\RequirePackage[2020-02-02]{latexrelease}

\documentclass[aps,preprintnumbers,showkeys,showpacs]{revtex4}
\usepackage{epsfig}
\usepackage{psfrag}
\usepackage{amsfonts}
\usepackage{graphicx}
\usepackage{dcolumn}
\usepackage{bm}
\usepackage{setspace}
\usepackage{multirow}
\usepackage{amsmath}
\usepackage{slashbox}

\begin{document}
\title{Testing the consistency of propagation between light and heavy cosmic ray nuclei}
\author{Yu Wang}
\author{Juan Wu}
\email{wu@cug.edu.cn}
\author{Wei-Cheng Long}
\affiliation{School of Mathematics and Physics, China University of Geosciences, Wuhan 430074, China}

\begin{abstract}
One of the fundamental issues in cosmic ray physics is to explain the nature of cosmic ray acceleration and propagation mechanisms. Thanks to the precise cosmic ray data measured by recent space experiments, we are able to investigate the cosmic ray acceleration and propagation models more comprehensively and reliably. In this paper, we combine the secondary-to-primary ratios and the primary spectra measured by PAMELA, AMS02, ACE-CRIS and Voyager-1 to constrain the cosmic ray source and transport parameters. The study shows that the $Z>2$ data yield a medium-energy diffusion slope $\delta_{2}\sim\left(0.42, 0.48\right)$ and a high-energy slope $\delta_{3}\sim\left(0.22, 0.34\right)$. The $Z\leq2$ species obtain a looser constraint on $\delta_{2}\sim\left(0.38, 0.47\right)$, but a tighter constraint on $\delta_{3}\sim\left(0.21, 0.30\right)$. The overlaps infer that the heavy and light particles can give compatible results at medium to high energies. Besides, both the light and heavy nuclei indicate a consistent diffusion slope variation $\Delta\delta_{H}$ around $200\sim300$~GV. At low energies, significant disagreements exist between the heavy and light elements. The B/C ratio requires a much larger diffusion slope shift $\Delta\delta_{L}$ around 4 GV or a stronger \textit{Alfv$\acute{e}$n velocity} $v_{A}$ than the low-mass data. This indicates that the heavy and light particles may suffer different low-energy transport behaviors in Galaxy. However, better understanding on the consistency/inconsistency between the heavy and light cosmic rays relies on more precise cross-sections, better constraints on correlations in systematic errors of data, more accurate estimation on Galaxy halo size and more robust description for Solar modulation during the reversal period of HMF.

\end{abstract}

\keywords{cosmic ray; propagation; acceleration}
\pacs{98.70Sa}
\maketitle

\section{Introduction}\label{sec:intro}
Benefiting from the development of detection technology, cosmic ray physics has entered a precise data-driven era. As we know, the secondary cosmic ray particles are produced by the primary particles interacting with the interstellar medium (ISM) when they transport in Galaxy. Therefore, the secondary-to-primary ratios reflect the propagation characteristics of cosmic rays. In previous theoretical studies, the cosmic ray propagation paradigm is commonly established by using the boron-to-carbon (B/C) ratio~\cite{Trotta2011, Niu2018, Yuan2019, Derome2019, Yuan2020}. But the importance of other secondary-to-primary ratios, especially the low-mass ones has been emphasized in literatures~\cite{Coste2012, Tomassetti2012, Wu2019, Weinrich2020}. Whether the light and heavy nuclei have the same experience in Galaxy is still under debate~\cite{Johannesson2016, Wu2019, Weinrich2020}. Recently, antiprotons are suggested to be an important probe to search dark matter signal~\cite{Cui2018, Reinert2018, Cuoco2019, Lin2019, Cholis2019, Jin2020}. However, if the light particles are not accelerated and propagated consistently with the heavy ones, uncertainties may exist in calculating antiproton background.

To improve our evaluation on cosmic ray physics, we use both the heavy and light nuclei data provided by PAMELA~\cite{Picozza2007}, AMS02~\cite{Aguilar2013}, ACE-CRIS~\cite{Stone1998} and Voyager-1~\cite{Stone1977} in our analysis. The heavy data used in this paper involve the carbon (C) spectrum and the corresponding secondary-to-primary ratios including the B/C ratio, the lithium-to-carbon (Li/C) ratio and the beryllium-to-carbon (Be/C) ratio. The low-mass data involve the protons (p) and helium (He) spectra, as well as the corresponding secondary-to-primary ratios such as the antiproton-to-proton ($\bar{\text{p}}$/p) ratio, the deuteron-to-helium 4 ($^2$H/$^4$He) ratio and the helium 3-to-helium 4 ($^3$He/$^4$He) ratio.

Since the AMS02 measurements cover a complex polarity reversal period in Heliospheric magnetic field (HMF) for which the solar effect is difficult to model, we only deal with the AMS02 data~\cite{Aguilar2015a, Aguilar2015b, Aguilar2016a, Aguilar2016b, Aguilar2017, Aguilar2018} (including p, He, C, $\bar{\text{p}}$/p, Li/C, Be/C, B/C) with energies larger than 20~GeV/n. On the contrary, no cut is performed on the PAMELA data~\cite{Adriani2010, Adriani2011, Adriani2014, Adriani2016} (including p, He, C, $\bar{\text{p}}$/p, $^2$H/$^4$He, $^3$He/$^4$He, B/C), since the PAMELA data was collected during a solar minimum period from 2006 to 2008, and the solar modulation during this period is simpler and easier to describe. We also employ the low-energy B/C and C data measured by ACE-CRIS~\footnote[1]{http://www.srl.caltech.edu/ACE/ASC/level2/lvl2DATA\_CRIS.html} during the same observational time of PAMELA. Besides, the energy spectra of interstellar cosmic rays observed by Voyager-1~\cite{Cummings2016} are included to help us further determine the validity of the studied models. With these high-precision data, we aim to study the cosmic ray acceleration and propagation mechanisms more comprehensively and to investigate whether the heavy and light nuclei yield compatible results.

\section{Parameter description}

Galactic cosmic rays are usually believed to be generated from supernova remnants and be accelerated at the expanding supernova shell via the diffusive shock acceleration. Then they are ejected into surrounding interstellar gas. For a certain type $i$ of particles, the source abundance can be written as:
\begin{equation}
\label{eq:source}
q_{i} =\left\{
\begin{aligned}
N_{ i} f (R) \rho^{-\nu_{1i}} & , & \rho< \rho_{bri} \\
N_{ i} f (R) \rho^{-\nu_{2i}} & , & \rho \geq \rho_{bri}
\end{aligned}
\right.
,
\end{equation}
where $R$ is the radial radius, $f\left(R\right)$ is the source spatial distribution in Galaxy, $N_{i}$ is the normalization abundance for the cosmic ray species $i$, and $\rho$ is the rigidity for the particle. Different injection indices $\nu_{1i}$ and $\nu_{2i}$ above and below a reference rigidity $\rho_{bri}$ are assumed. In our previous work~\cite{Wu2019}, we made an assumption that the injection indices for helium nuclei were correlated with those for protons. But here we allow independent injection indices for different species in order to find out the real patterns of source parameters. In this work, we use the cosmic ray propagation software GALPROP v54~\footnote{https://gitlab.mpcdf.mpg.de/aws/galprop}~\cite{Moskalenko1998, Strong1998, Strong2000, Strong2015} to calculate the cosmic ray interstellar spectra. The normalization abundances of all other primary species are set relative to the source abundance of protons $N_{\text{p}}$. Therefore, for species such as He and C, their normalization abundances can be written as $N_{i}=X_{i}N_{\text{p}}$, where $X_{i}$ is the ratio of the normalized abundance of a certain species $i$ of particles to that of protons. The source radial distribution is assumed to be $f (R) = (R /R_{\odot})^{\alpha} e^{-\beta (R-R_{\odot}} )$, where $R_{\odot} = 8.5$~kpc. In this study, we choose $\alpha = 0.475$ and $\beta = 1.166$.

After entering the interstellar space, the cosmic rays are influenced by the irregular magnetic field and are scattered randomly in Galaxy. The diffusion coefficient is assumed to be:
\begin{equation}
\label{eq:diff_coeffieient}
D_{xx}=\left\{
\begin{aligned}
& D_{0}\beta^{\eta} \left(\frac{\rho}{\rho_{0}}\right)^{\delta}  , & \rho< \rho_{1} \\
& D_{0}\beta^{\eta} \left(\frac{\rho_{1}}{\rho_{0}}\right)^{\delta} \left(\frac{\rho}{\rho_{1}}\right)^{\delta_{3}}  , & \rho\geq \rho_{1}
\end{aligned}
\right.
,
\end{equation}
where $\beta=\upsilon/c$ is the particle velocity, $D_{0}$ is the normalization of diffusion coefficient at a reference rigidity $\rho_{0}$,  $\eta$ is a low-energy dependence factor which may be related to the magnetohydrodynamic turbulence dissipation effect~\cite{Ptuskin2006}, $\delta$ is the diffusion slope at rigidities below $\rho_{1}$. At rigidities above $\rho_{1}$, we introduce a high-energy diffusion slope $\delta_{3}$ in light of the hardening at a few hundreds GV observed in primary fluxes~\cite{Aguilar2015a, Aguilar2015b,Aguilar2017,Adriani2011} and a stronger hardening for secondaries~\cite{Aguilar2018}. Moreover, for the pure diffusion model, $\delta$ needs to take different values $\delta_{1}$ and $\delta_{2}$ below and above $\rho_{0}$~\cite{Wu2019, Genolini2019, Vittino2019, Weinrich2020, Korsmeier2021}.

In addition to diffusion in position space, diffusion may also happen in momentum space due to the interaction between cosmic rays and the magnetic turbulence. As a result, cosmic ray particles may be reaccelerated. The associated diffusion coefficient in momentum space $D_{pp}$ is correlated to the spatial diffusion coefficient $D_{xx}$ as follows:
\begin{equation}
D_{pp}=\frac{4v_{A}^{2}p^{2}}{3\delta\left(4-\delta^{2}\right)\left(4-\delta\right)D_{xx}},
\end{equation}
where $v_{A}$ is the \textit{Alfv\'{e}n velocity}, corresponding to the turbulence velocity in the hydrodynamical plasma. The magnitude of $v_{A}$ represents the strength of reacceleration effect. For the diffusion reacceleration model, there is no need for a low-energy break on the diffusion slope, i.e. $\delta_{2}=\delta_{1}$. Besides, cosmic rays may also suffer a convection process that transports particles from Galactic disk to Galactic halo. Different assumptions were made in literatures. For example, a constant convection velocity was assumed in \cite{Reinert2018, Cuoco2019}, and a linear velocity was adopted in \cite{Cholis2019, Boschini2020}. In this paper, we do not take into account the convection process, but will further study this mechanism in our future work.

After cosmic rays enter the solar system from interstellar space, they are modulated by the solar wind. In this paper, we use force-field approximation~\cite{Gleeson1968} to describe the heliospheric modulation. This model can generally describe the periodic data collected during the solar minimum period~\cite{Corti2019}. For a nucleus with charge $Z$, mass $m$ and atomic number $A$, the modulated cosmic ray energy spectrum at the top of atmosphere $J_{\text{TOA}}$ and the unmodulated interstellar flux $J_{\text{IS}}$ are related as:
\begin{equation}
J_{\text{TOA}}\left( E \right)=\frac{\left( E+m \right)^2-m^2}{\left(E+m+\frac{\left|Z\right|}{A}\phi\right)^2-m^2}J_{\text{IS}}\left( E+\frac{\left|Z\right|}{A}\phi \right),
\end{equation}
where $E$ is the kinetic energy of the nucleus and $\phi$ is the modulation potential.

To summarize, those parameters describing the cosmic ray acceleration and propagation mechanisms include the source parameters, the propagation parameters and the solar modulation parameter. The source parameters involve $\nu_{1\text{p}}$, $\nu_{2\text{p}}$, $\rho_{br\text{p}}$, $N_{\text{p}}$, $\nu_{1\text{He}}$, $\nu_{2\text{He}}$, $\rho_{br\text{He}}$, $X_{\text{He}}$ for the light nuclei with charge number $Z\leq2$, and $\nu_{1\text{C}}$, $\nu_{2\text{C}}$, $\rho_{br\text{C}}$, $X_{\text{C}}$ for the heavy nuclei with charge number $Z>2$. The propagation parameters involve $D_{0}$, $\delta_{1}$, $\delta_{2}$, $\delta_{3}$, $\rho_{0}$, $\rho_{1}$, $\eta$ and $v_{A}$. The solar modulation parameter contains only one single parameter $\phi$. Since the degeneracy between $D_{0}$ and the halo size of Galaxy $z_{h}$ can only be broken by radioactive species, we choose $z_{h}=4$~kpc to
be consistent with earlier studies~\cite{Strong2001, Boschini2017, Wu2019, Boschini2020}. But it should be aware that the error estimation on other parameters can be underestimated since $z_{h}$ is fixed. It is also important to note that the GALPROP resolution parameters will influence the accuracy of the calculation. To compromise between calculation speed and accuracy, the GALPROP spatial, energy and time resolution parameters are set as the values given in Table~\ref{tab:GalRes}.

\begin{table*}[!htbp]
\begin{spacing}{1.2}
\begin{center}
\small
\begin{tabular}{c | c | c}
\hline \hline
Resolution parameter & Explanation & Value  \\
\hline
$dr$   & radial grid size & 1.0~kpc\\
$dz$  & height grid size  & 0.2~kpc \\
$Ekin$\_$factor$  & kinetic energy spacing on a logarithmic scale & 1.3\\
$timestep$\_$factor$  & scaling factor for timestep reducing & 0.25 \\
$start$\_$timestep$  & initial timestep & $10^{9}$~s \\
$end$\_$timestep$  & final timestep & $10^{2}$~s \\
$timestep$\_$repeat$ & timestep repetitions for each timestep\_factor & 20 \\
\hline
\end{tabular}
\end{center}
\end{spacing}
\caption[GALPROP resolution]{\footnotesize The numerical scheme parameters of GALPROP adopted in this study.}\label{tab:GalRes}
\end{table*}

\section{Results}
To analyse the experimental data, the $\chi^{2}$ minimization method is used in this paper. Compared with the Bayesian analysis, this method consumes less computation time and can efficiently estimate the best-fit parameters. It also gives the goodness of fit of each model, i.e. the minimum $\chi^{2}$ value. Specifically, we interface the minimization library MINUIT~\cite{James1975} with GALPROP to implement the parameter estimation. For each given parameter, the MINUIT processor MINOS is utilized to reliably calculate its asymmetric errors. The positive and negative MINOS errors are defined as the changes in the value of that parameter which causes the minimum $\chi^{2}$ value to increase by 1. To achieve accurate estimations on best-fit parameters and their errors, nearly O(10$^4$) GALPROP runs are required. For heavy nuclei, the nuclear network starts at $^{28}$Si and the heavy element scan takes nearly 1.4 CPU minutes per run. For light elements, the nuclear chain starts at $^{4}$He and the scan requires approximately 22 seconds per run.

Fittings are performed by using the $Z>2$ and $Z\leq2$ nuclei separately. To accurate describe the secondary component of antiprotons, we employ the updated cross-section data provided by~\cite{Korsmeier2018} and embed a code from~\cite{Lin2017, Lin2021} in Galprop to calculate the antiproton productions. For $^2$H and $^3$He, based on those cross-sections derived from~\cite{Coste2012}, we modify the corresponding data in GALPROP file `eval\_iso\_cs.dat' to better determine the productions of $^2$H and $^3$He. Some literatures found a possible existence of primary Li~\cite{Boschini2020}, or uncertainties on Li production cross-section~\cite{Weinrich2020,Korsmeier2021}. To account for these effects, we introduce an scaling factor on the Li production, i.e. $S_{\text{Li}}$. Two different propagation frameworks are studied in this paper: (1) PDbr model: the plain diffusion model with a low-energy break around a few~GV in the diffusion slope, i.e. $\delta_{1}\neq \delta_{2}$; (2) DR model: the diffusion-reacceleration model without a low-energy break in the diffusion slope, i.e. $\delta_{1}= \delta_{2}$.

\subsection{Fit to the heavy elements}\label{sec:heavyelements}

We first investigate the PDbr and DR models by utilizing a data set combination with the B/C and C data. The corresponding models are defined as the reference models for heavy particles and are suffixed with ``-H0''. Since Li and Be, like B, are the secondaries produced by C interacting with ISM, we further add the accurate AMS02 Li/C and Be/C data in the analysis to check whether they give compatible results with those derived from only the B/C and C data. When the (Li, Be, B)/C and C data are all included to run the fitting, the corresponding models are suffixed with ``-H''. Since the normalization abundance of C is calculated according to the abundance of protons, the injection parameters for protons may have impact on the source term of C. Therefore, when we analyse C and its secondaries, the proton source parameters are fixed at the best-fit values derived from all the light particles, which will be detailed later in Table~\ref{tab:Z2results}. By fitting the heavy elements, the estimated source and propagation parameters and the minimized $\chi^2$ value for each model are shown in Table \ref{tab:BCresults}.

\begin{table*}[!htbp]
\begin{spacing}{0.8}
\begin{center}
\small
\begin{tabular}{l  c  c  c  c }
\hline \hline
Parameter & PDbr-H0& DR-H0 & PDbr-H & DR-H \\
\hline
$D_{0}$ $(10^{28}$cm$^{2}$\,s$^{-1})$ & $3.2\pm^{0.4}_{0.3}$ & $2.18\pm^{0.15}_{0.14}$ & $3.34\pm^{0.25}_{0.22}$ & $3.00\pm^{0.23}_{0.22}$ \\
$\delta_{1}$ & $-1.6\pm0.4$ &$0.435\pm^{0.019}_{0.018}$&$-1.4\pm0.4$ &$0.446\pm0.014$\\
$\delta_{2}$&$0.462\pm0.016$&[$=\delta_{1}$]&$0.472\pm^{0.013}_{0.012}$ &[$=\delta_{1}$]\\
$\delta_{3}$&$0.31\pm^{0.03}_{0.04}$&$0.26\pm^{0.03}_{0.04}$&$0.310\pm^{0.026}_{0.027}$&$0.29\pm^{0.03}_{0.04}$\\
$\rho_{0}$ (GV)&$3.86\pm^{0.19}_{0.15}$&[4]&$3.91\pm^{0.20}_{0.15}$ &[4]\\
$\rho_{1}$ ($10^{2}$GV)&$2.2\pm^{0.4}_{0.3}$&$2.2\pm^{0.5}_{0.3}$&$2.1\pm0.4$ &$2.3\pm^{0.5}_{0.4}$\\
$\eta$& $2.5\pm0.7$ & $0.04\pm^{0.19}_{0.20}$ & $2.2\pm^{0.7}_{0.6}$ &$-0.27\pm0.17$ \\
$v_{A}$ (km\,s$^{-1}$)&---&$17.6\pm1.0$ &---&$17.2\pm^{1.8}_{1.9}$\\
$\nu_{1\text{C}}$& $0.32\pm^{0.14}_{0.15}$ & $0.68\pm^{0.12}_{0.20}$ &$0.36\pm^{0.11}_{0.12}$ & $1.42\pm0.04$\\
$\nu_{2\text{C}}$& $2.338\pm0.012$ & $2.412\pm0.010$ & $2.328\pm0.010$ &$2.353\pm^{0.011}_{0.012}$ \\
$\rho_{br\text{C}}$ (GV)& $1.22\pm^{0.06}_{0.05}$ & $1.51\pm^{0.09}_{0.12}$ &$1.22\pm0.04$&$2.47\pm0.10$\\
$X_{\text{C}}$ ($10^{-3}$) & $3.3\pm^{0.5}_{0.4}$ & $23\pm^{5}_{3}$ &$3.12\pm^{0.27}_{0.26}$&$5.0\pm^{0.8}_{0.7}$  \\
$S_{\text{Li}}$  & [1] & [1] &$1.234\pm0.013$&$1.243\pm0.013$  \\
$\phi$ (GV)& $0.442\pm0.014$ &$0.514\pm^{0.013}_{0.012}$ &$0.443\pm0.012$&$0.409\pm0.013$ \\
\hline
$\chi^{2}/$d.o.f&0.92&0.91& 0.86 &1.45 \\
\hline
\end{tabular}
\end{center}
\end{spacing}
\caption[The results from heavy particles]{\footnotesize The best-fit parameters for PDbr-H0, DR-H0, PDbr-H and DR-H models constrained by the $Z>2$ data. The fixed parameters appear in square brackets. }\label{tab:BCresults}
\end{table*}

\begin{figure}[!htbp]
\centering
\includegraphics[width=8.5cm]{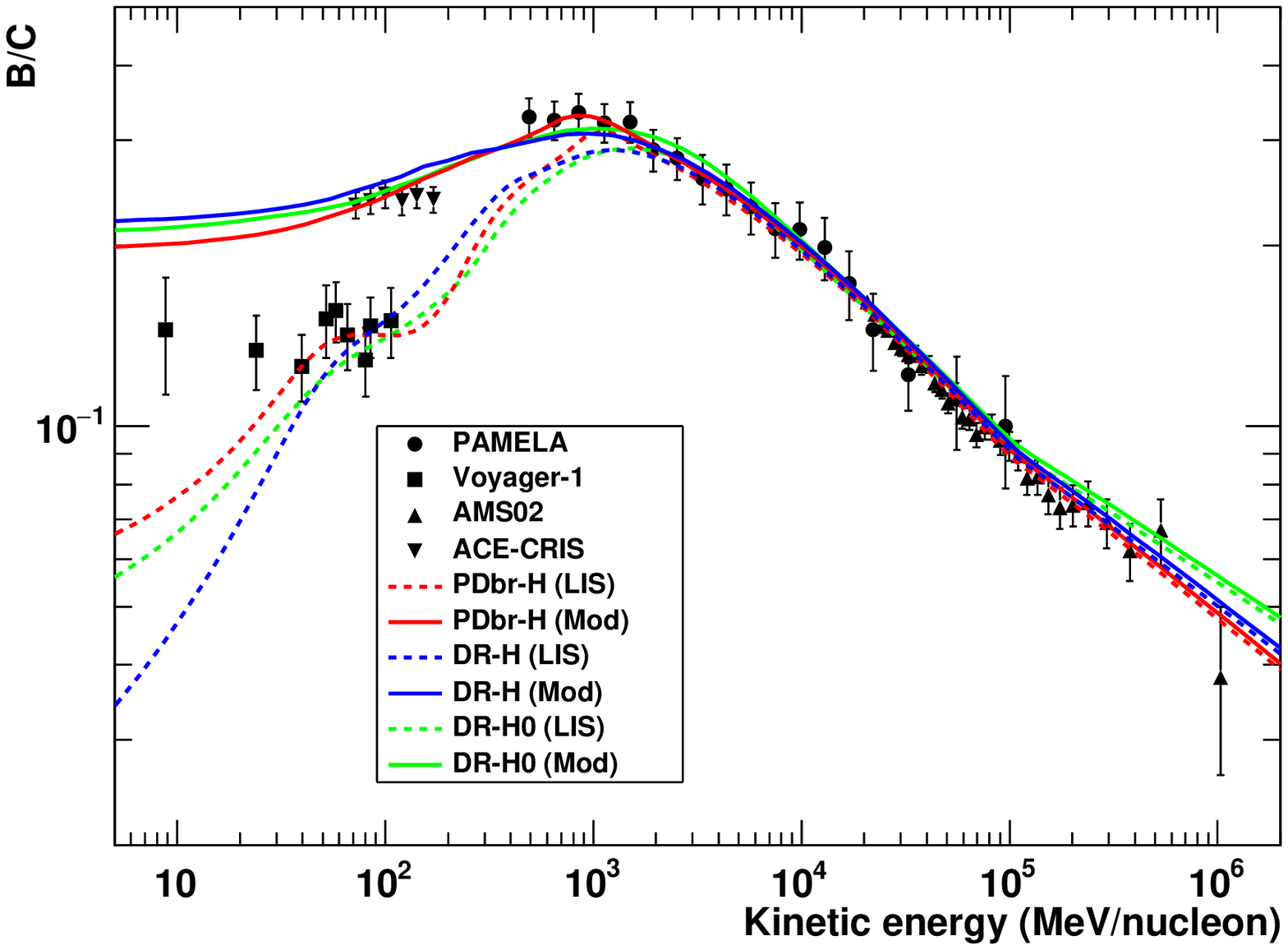}
\includegraphics[width=8.5cm]{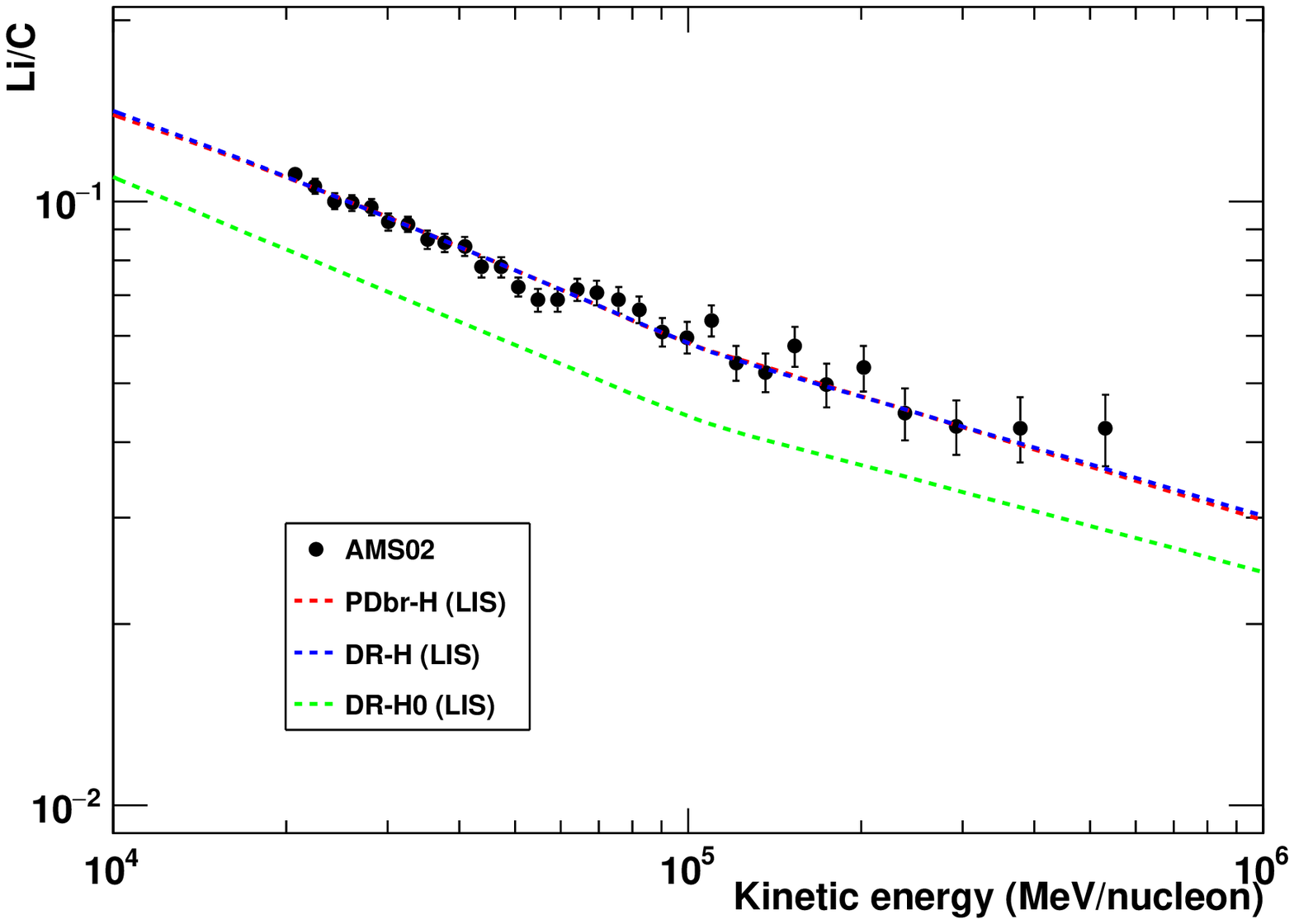}
\includegraphics[width=8.5cm]{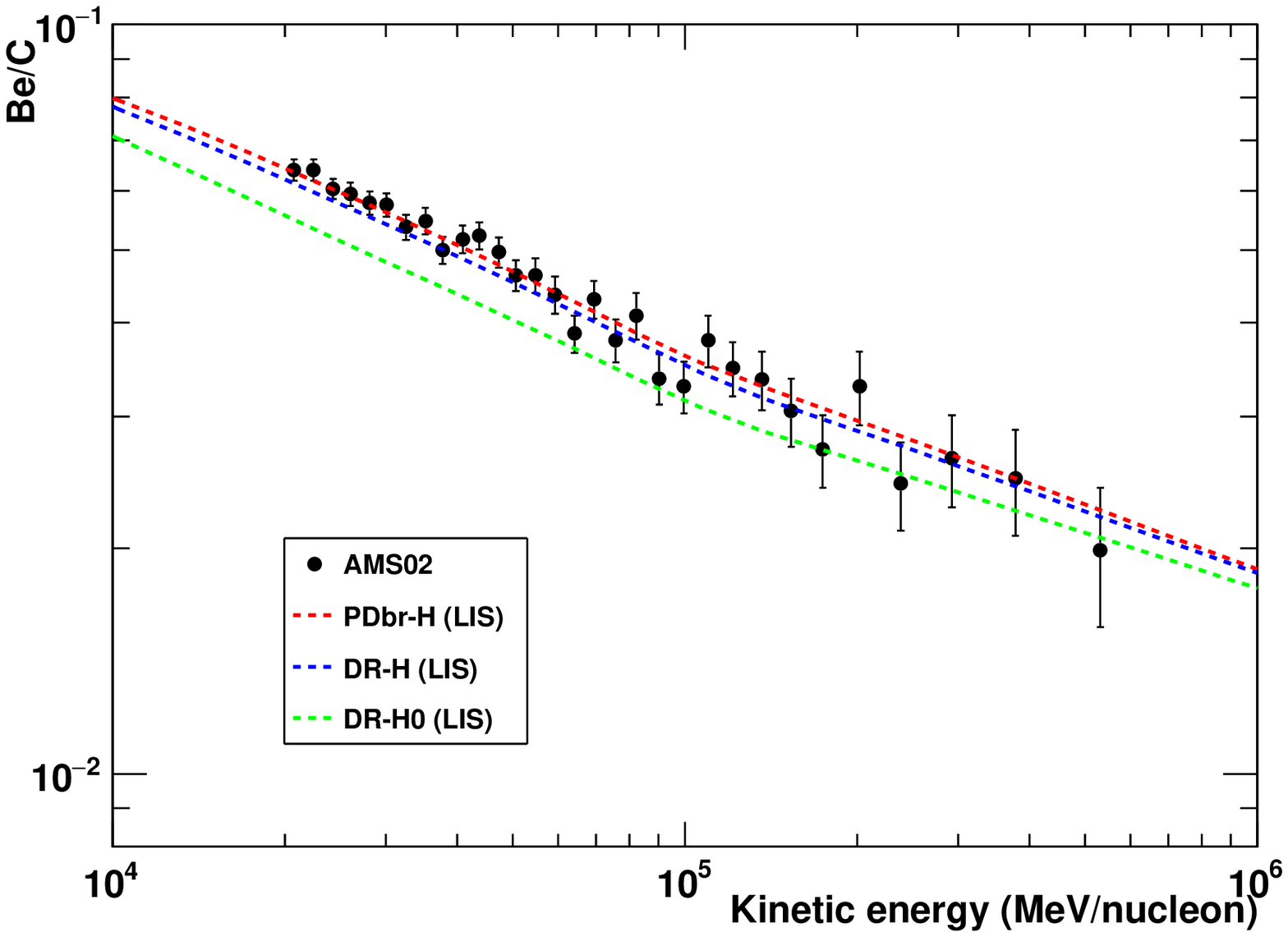}
\includegraphics[width=8.5cm]{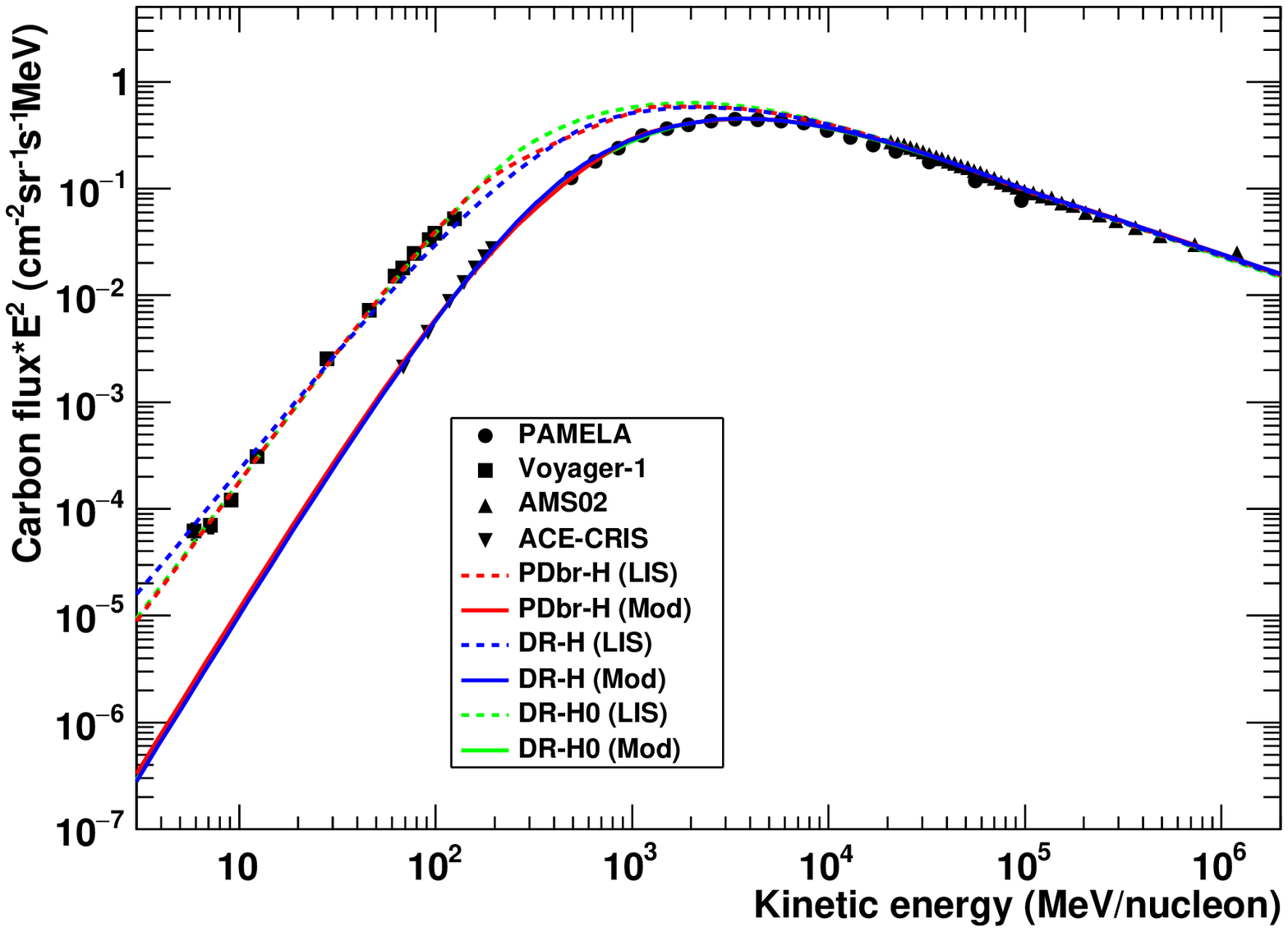}
\caption[The heavy S/P ratios and the fluxes in the $\chi^{2}$ study]{\footnotesize The B/C, Li/C, Be/C ratios and the carbon flux for the best-fit parameters of PDbr-H, DR-H and DR-H0 models as listed in Table~\ref{tab:BCresults}. The calculation from PDbr-H0 model cannot be distinguished with that from PDbr-H and is not shown here. The solid (dashed) lines represent the interstellar (modulated) spectra and ratio. Data points are the measurements from PAMELA, AMS02, ACE-CRIS and Voyager-1.}\label{fig:LiBeC}
\end{figure}

It is found that the best-fit parameters for PDbr-H model are consistent with those for PDbr-H0 model, with slightly improved accuracies. The best-fit propagation parameters for DR-H model also agree well with those for DR-H0 model. However, the best-fit source and solar modulation parameters for DR-H model, i.e. $\nu_{1\text{C}}$, $\nu_{2\text{C}}$, $X_{\text{C}}$ and $\phi$, differ significantly with those for DR-H0 model. Moveover, as exhibited in Fig~\ref{fig:LiBeC}, the theoretical calculation of Li/C for DR-H0 model disagrees dramatically with other models as well as the experimental data. This discrepancy on Li may be caused by different values of $S_{\text{Li}}$ adopted in the calculation. But the prediction of Be/C from DR-H0 model shows distinct disagreements with the AMS02 data below 60~GeV/n. It indicate that using only the B/C and C data may not provide reliable constraints on the source and transport mechanisms for heavy particles. In the following paragraphs, we concentrate on discussing PDbr-H and DR-H models.

For PDbr-H and DR-H models, the diffusion spectral index $\delta_{2}$ are well constrained in $\left(0.43, 0.48\right)$. The $\Delta\delta_{H}=\delta_{3}-\delta_{2}$ values estimated in PDbr-H and DR-H models are $-0.16\pm0.03$ and $-0.16\pm^{0.04}_{0.05}$, respectively. These results indicate that a consistent change in the diffusion slope around $200\sim300$~GV exists in both models, which may be responsible for the hardening in primary and secondary cosmic ray spectra above a few hundreds~GV. Though the high-energy behaviors are explicit, the phenomena at low energies are model-dependent. For example, the estimated $\eta$ is much smaller when a reacceleration process is considered. Since the Li/C and Be/C data employed in the analysis are those with energies larger than 20~GeV/n, the low-energy propagation parameters are mainly constrained by the B/C data. To explain the B/C peak around 1~GeV/n, either a diffusion slope variation $\Delta\delta_{L}=\delta_{2}-\delta_{1}\sim1.9\pm0.5$ around 4~GV or a \textit{Alfv$\acute{e}$n velocity} $v_{A}\sim17.2\pm^{1.8}_{1.9}$~km\,s$^{-1}$ is required. However, it can be found in Fig~\ref{fig:LiBeC}, both models expect lower ratios than the Voyager-1 data below 20~MeV/n, as already emphasized in \cite{Cummings2016}. The best-fit values of $S_{\text{Li}}$ are found to be $1.234\pm0.013$ and $1.243\pm0.013$ in PDbr-H and DR-H models. These results can either been seen as a possible signal of primary Li, or a hint for an inaccurate cross-section normalization on Li. Nevertheless, both PDbr-H and DR-H models can generally reproduce all the $Z>2$ data.

\subsection{Fit to the light elements}\label{sec:lightelements}
To better understand the cosmic ray source and transport phenomena for light elements, we further implement a $\chi^{2}$ analysis on the Z$\leq$2 species. Similar with the treatment on heavy particles, we first examine the PDbr and DR models by employing a commonly-used data set combination with the $\bar{\text{p}}$/p, p and He data. The corresponding models are suffixed with ``-L0''. Then we include the $^2$H/$^4$He and $^3$He/$^4$He data in the analyses. The corresponding models are suffixed with ``-L''. The estimated best-fit parameters and the $\chi^{2}$ value for each model are presented in Table~\ref{tab:Z2results}.

\begin{table*}[!htbp]
\begin{spacing}{0.7}
\begin{center}
\small
\begin{tabular}{l  c  c  c  c }
\hline \hline
Parameter & PDbr-L0& DR-L0 & PDbr-L & DR-L \\
\hline
$D_{0} $ ($10^{28}$cm$^{2}$\,s$^{-1}$) & $3.96\pm^{0.13}_{0.12}$ & $3.47\pm0.08$ & $3.76\pm^{0.07}_{0.08}$ & $4.10\pm0.08$ \\
$\delta_{1}$ & $-0.09\pm0.05$ & $0.462\pm0.010$& $-0.23\pm^{0.04}_{0.05}$ & $0.386\pm0.009$ \\
$\delta_{2}$ &$0.410\pm0.013$&[$=\delta_{1}$]&$0.409\pm^{0.009}_{0.008}$&[$=\delta_{1}$]\\
$\delta_{3}$ &$0.256\pm^{0.023}_{0.020}$&$0.276\pm^{0.023}_{0.027}$&$0.257\pm^{0.020}_{0.019}$&$0.226\pm^{0.022}_{0.017}$\\
$\rho_{0}$ (GV)&$4.62\pm^{0.25}_{0.17}$&[4]&$4.07\pm^{0.12}_{0.16}$&[4]\\
$\rho_{1}$ ($10^{2}$GV)&$3.2\pm^{0.3}_{0.4}$&$4.2\pm0.4$&$3.3\pm^{0.3}_{0.4}$&$4.0\pm^{0.3}_{0.6}$\\
$\eta$ &$1.06\pm0.14$ &$-0.03\pm0.08$ &$1.10\pm0.10$&$-0.20\pm^{0.06}_{0.07}$\\
$v_{A}$ (km\,s$^{-1}$)&---&$15.8\pm^{0.7}_{0.8}$&---&$12.6\pm^{1.0}_{1.1}$\\
$\nu_{1\text{p}}$  & $1.543\pm0.026$ & $1.910\pm0.015$ & $1.47\pm^{0.04}_{0.05}$ & $1.781\pm0.016$\\
$\nu_{2\text{p}}$  & $2.418\pm0.013$ & $2.344\pm0.010$ & $2.417\pm^{0.008}_{0.010}$ & $2.419\pm0.008$ \\
$\rho_{br\text{p}}$ (GV) & $1.74\pm^{0.08}_{0.07}$ & $5.9\pm0.4$ & $1.40\pm^{0.09}_{0.13}$ & $3.18\pm^{0.11}_{0.10}$\\
$\nu_{1\text{He}}$  & $1.439\pm0.021$ & $1.593\pm0.019$ & $1.455\pm^{0.023}_{0.019}$ & $1.492\pm^{0.019}_{0.018}$ \\
$\nu_{2\text{He}}$  & $2.357\pm0.012$ & $2.280\pm0.009$ & $2.355\pm^{0.008}_{0.009}$ & $2.365\pm0.007$\\
$\rho_{br\text{He}}$ (GV) & $2.43\pm0.06$ &$2.45\pm^{0.12}_{0.10}$& $2.361\pm^{0.022}_{0.013}$ &$2.36\pm^{0.04}_{0.02}$ \\
$N_{\text{p}}$ ($10^{-9}$cm$^{-2}$\,sr$^{-1}$\,s$^{-1}$\,MeV$^{-1}$)& $4.317\pm0.012$&$4.313\pm0.012$ &$4.318\pm0.012$&$4.323\pm^{0.011}_{0.012}$\\
$X_{\text{He}}$ & $0.032\pm^{0.004}_{0.003}$ & $0.60\pm^{0.14}_{0.11}$& $0.020\pm^{0.004}_{0.005}$ & $0.151\pm0.011$ \\
$\phi$ (GV)& $0.423\pm0.007$ &$0.442\pm^{0.007}_{0.008}$ & $0.424\pm^{0.007}_{0.008}$ &$0.441\pm^{0.006}_{0.007}$ \\
\hline
$\chi^{2}$/d.o.f &1.76 &1.90 &1.67&2.15\\
\hline
\end{tabular}
\end{center}
\end{spacing}
\caption[The results from light particles]{\footnotesize The best-fit parameters for PDbr-L0, DR-L0, PDbr-L and DR-L models constrained by the Z$\leq$2 data. The fixed parameters appear in square brackets. }\label{tab:Z2results}
\end{table*}

\begin{figure}[!htbp]
\centering
\includegraphics[width=8.5cm]{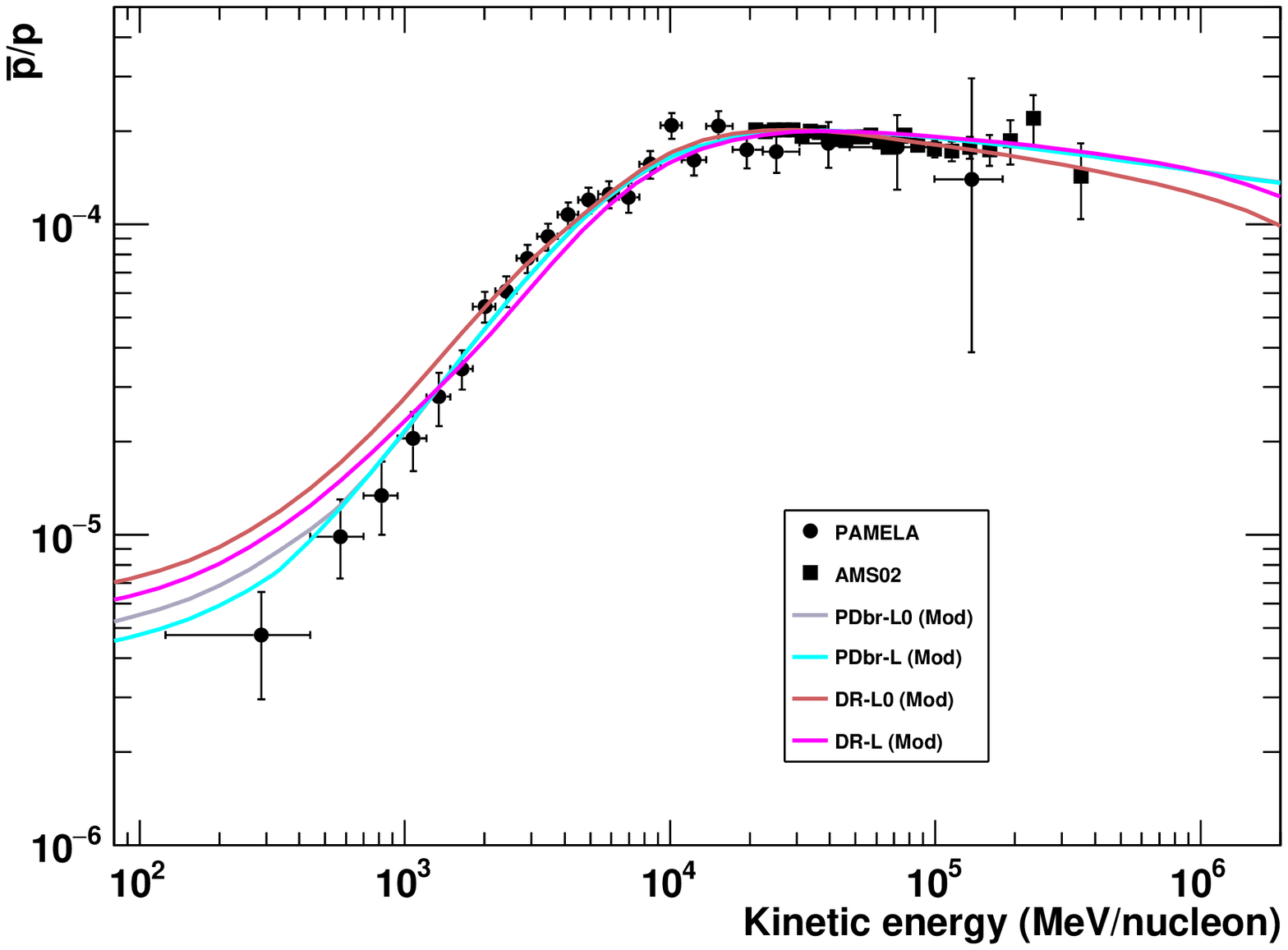}
\includegraphics[width=8.5cm]{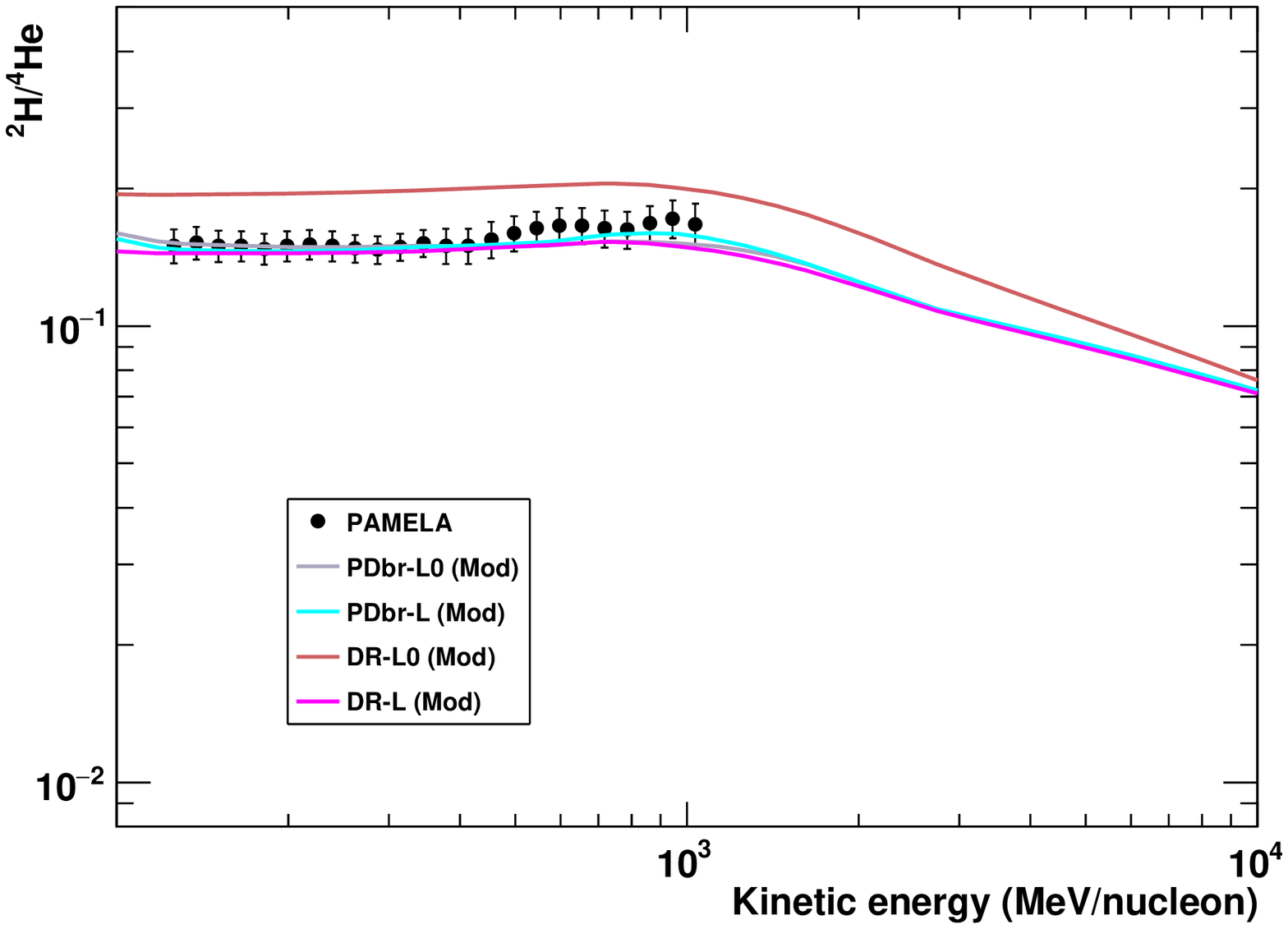}
\includegraphics[width=8.5cm]{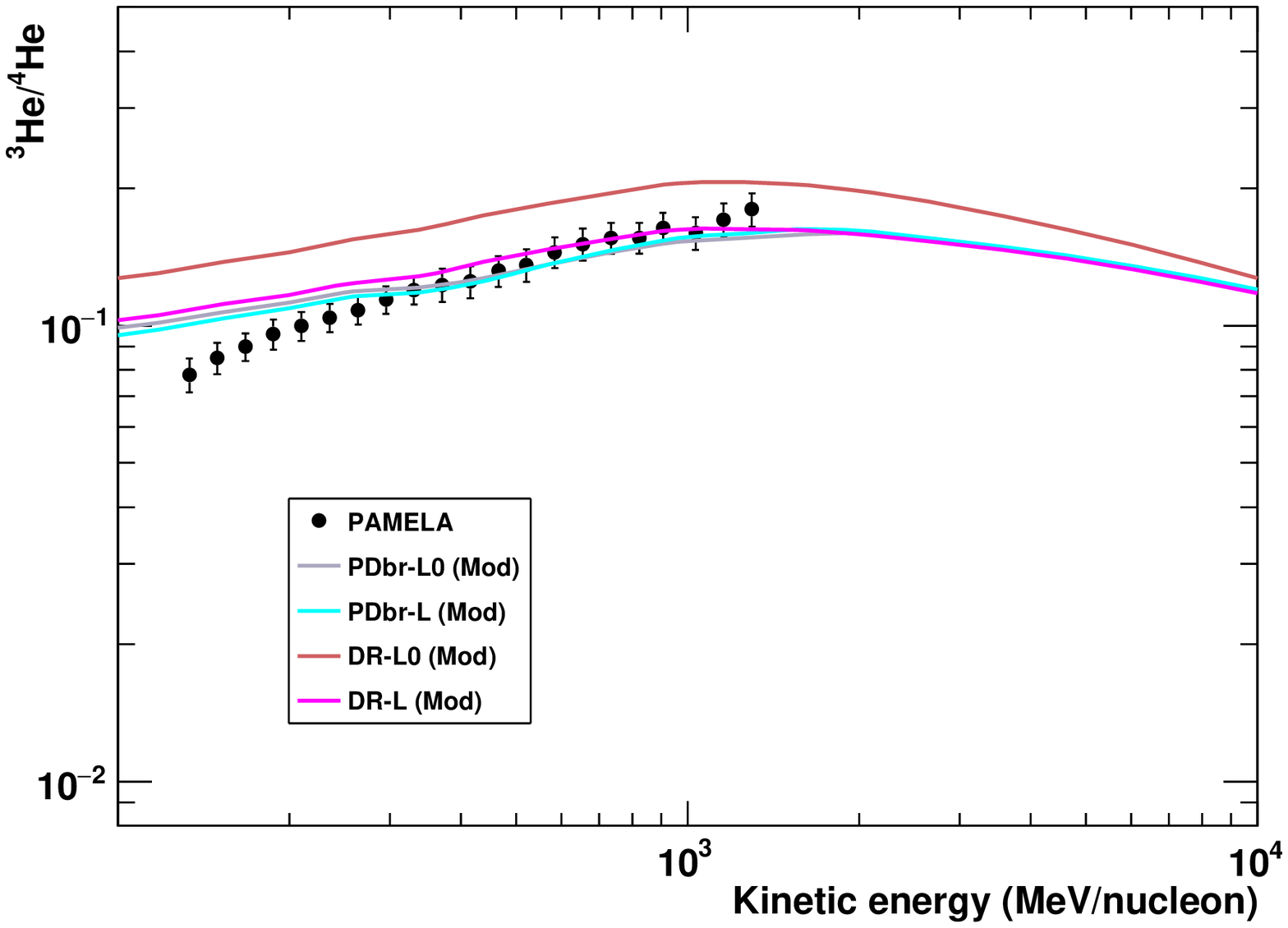}
\includegraphics[width=8.5cm]{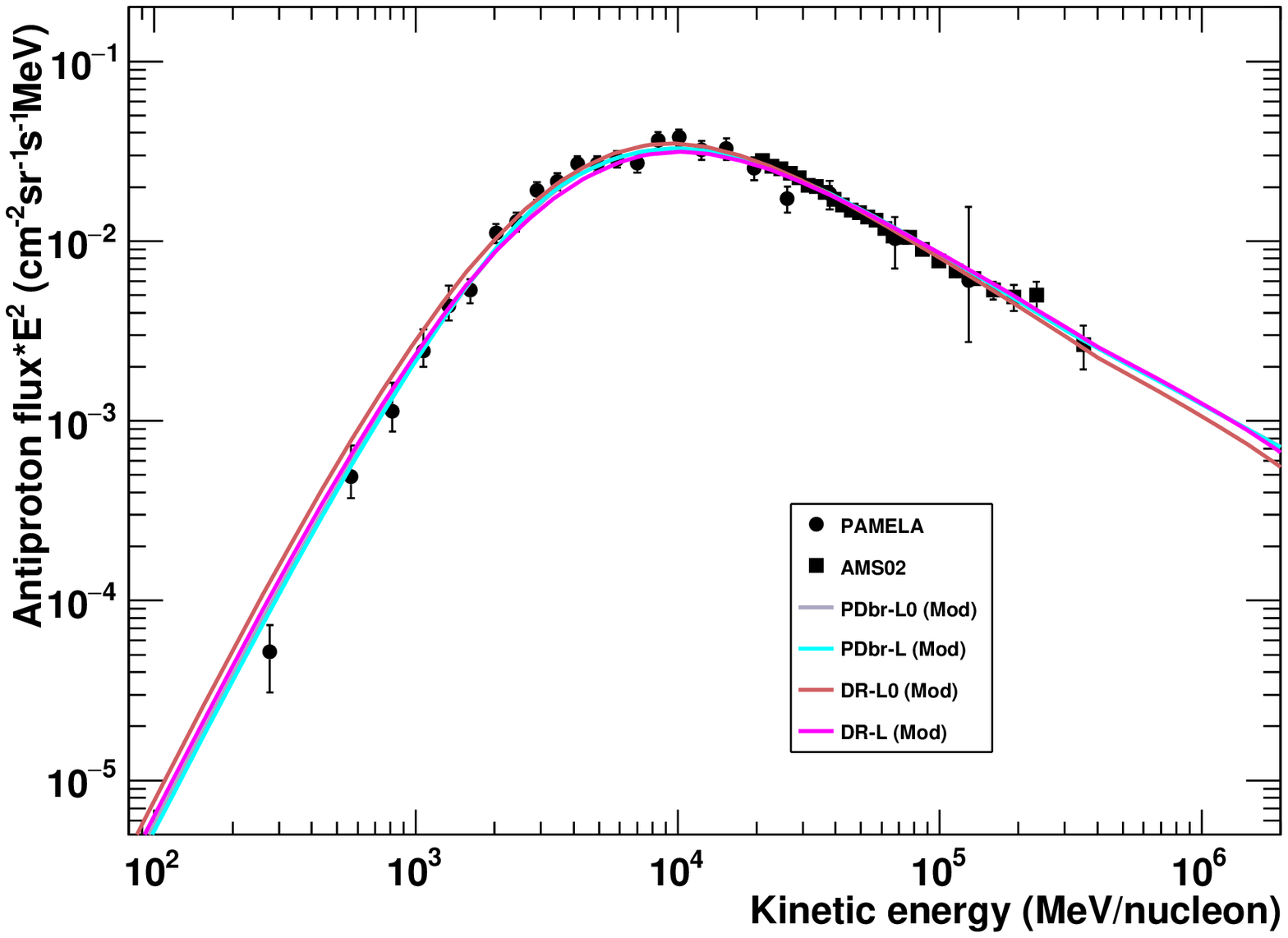}
\includegraphics[width=8.5cm]{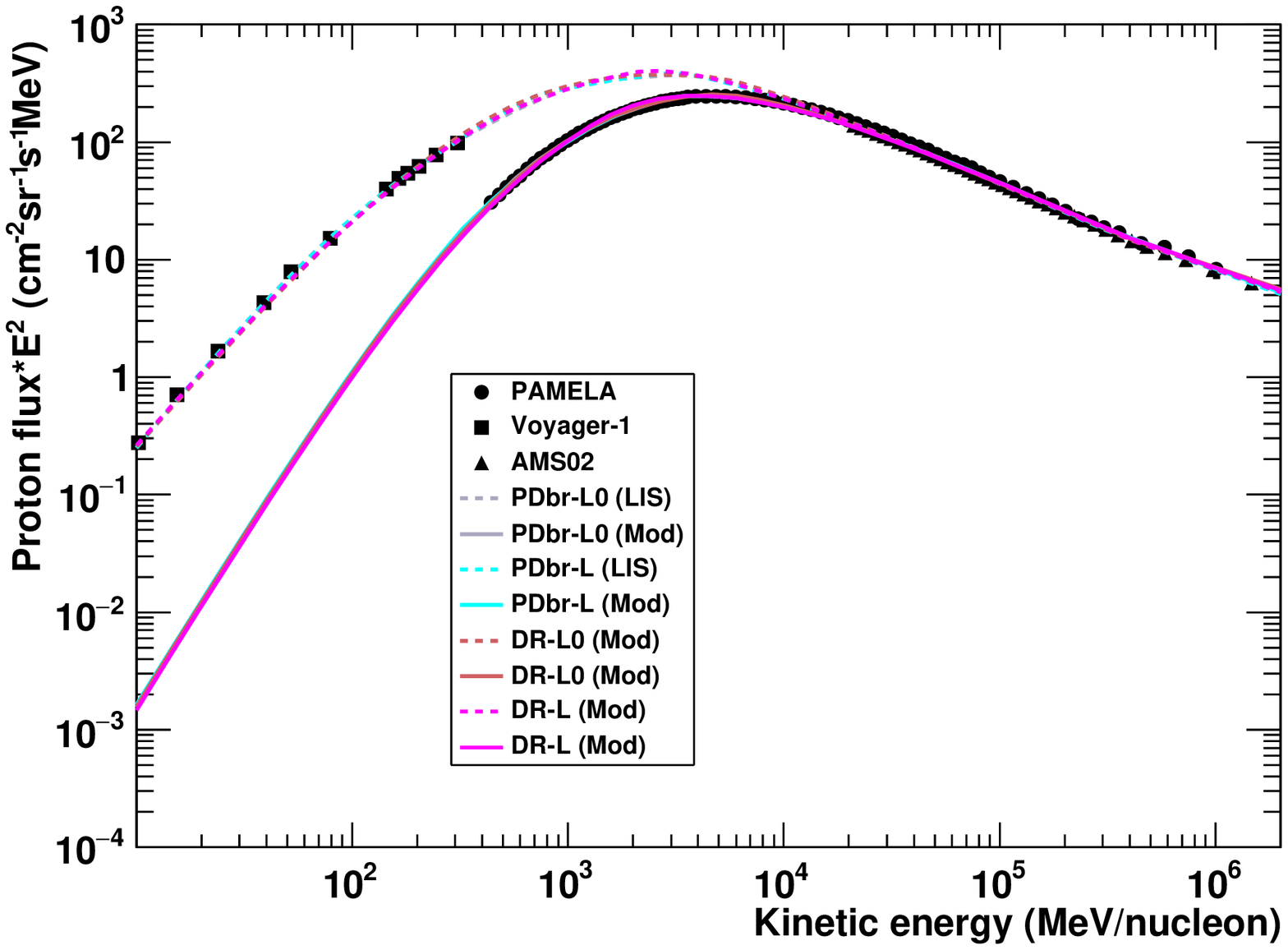}
\includegraphics[width=8.5cm]{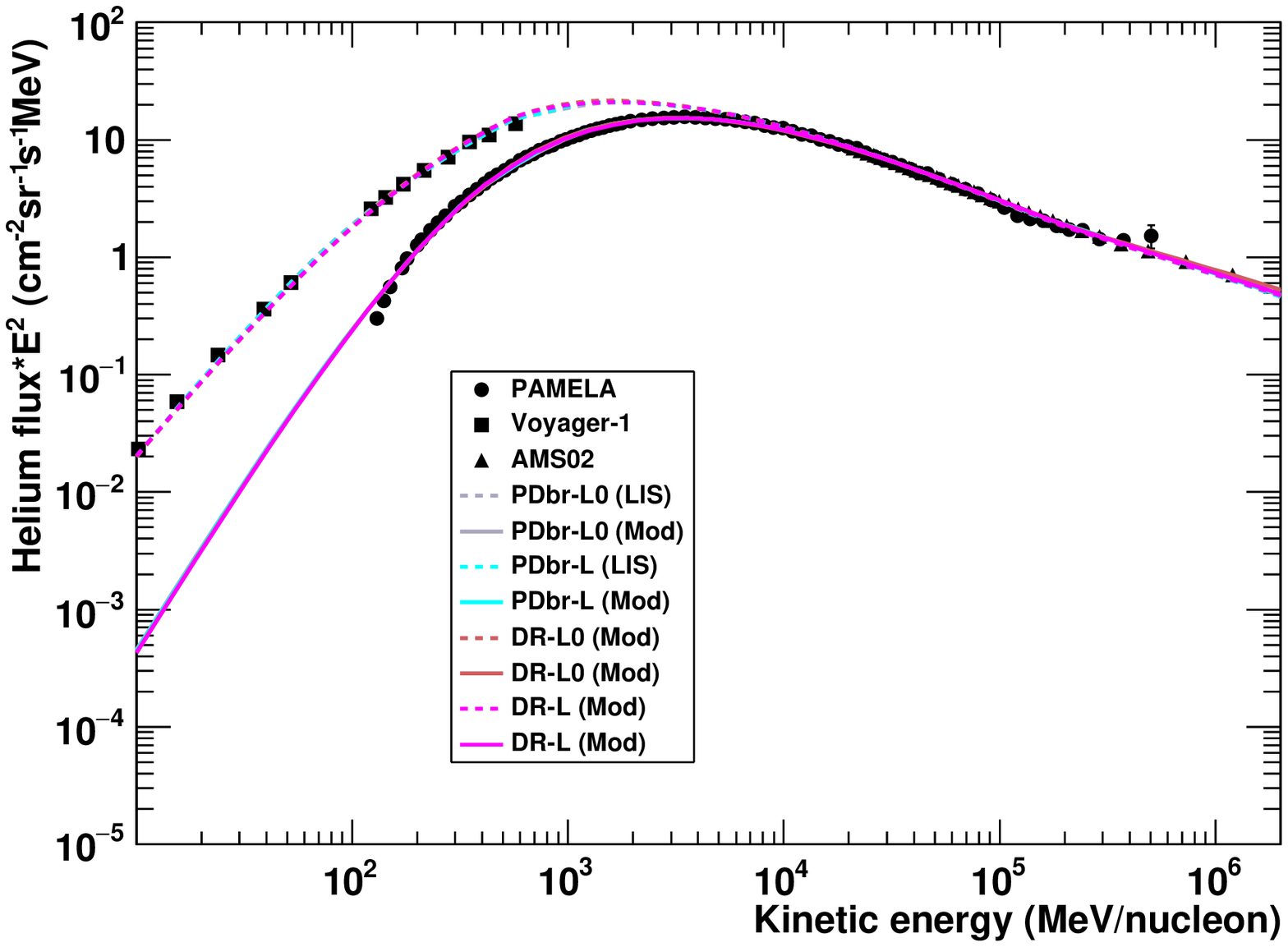}
\caption[The low mass S/P ratios and the fluxes in the $\chi^{2}$ study]{\footnotesize The $\bar{\text{p}}$/p, $^{2}$H/$^{4}$He, $^{3}$He/$^{4}$He ratios and the antiproton, proton, helium fluxes for the best-fit parameters of PDbr-L0, PDbr-L, DR-L0 and DR-L models as listed in Table~\ref{tab:Z2results}. The solid (dashed) lines represent the interstellar (modulated) spectra and ratio. Data points are the measurements from PAMELA, AMS02 and Voyager-1.}\label{fig:Z2}
\end{figure}

The best-fit values of $\delta_{2}$ are $0.410\pm0.013$ for PDbr-L0 model and $0.462\pm0.010$ for DR-L0 model. These two values generally agree with those derived from heavy particles, which indicates that the $\bar{\text{p}}$/p data can yield compatible results on $\delta_{2}$ with the B/C data. By adding the $^2$H/$^4$He and $^3$He/$^4$He ratios in the fitting, PDbr-L model gives consistent results with PDbr-L0 model. However, it is found that DR-L model yields inconsistent parameters compared with DR-L0 model. For example, the $\delta_{2}$ value is varied prominently from $0.462\pm0.010$ in DR-L0 model to $0.386\pm0.011$ in DR-L model. This is because DR-L0 model can not reproduce the PAMELA $^2$H/$^4$He, $^3$He/$^4$He data below 1~GeV/n, as exhibited in Fig~\ref{fig:Z2}. Therefore, an inclusion of the $^2$H/$^4$He and $^3$He/$^4$He data in the fitting leads to different estimations on source and transport parameters in DR-L model. In the following paragraphs, we focus on discussing PDbr-L and DR-L models.

Under the same configuration, i.e. the PDbr configuration or the DR configuration, the value of $\delta_{2}$ determined by $\bar{\text{p}}$, $^{2}$H, $^{3}$He, p and He, is about 0.06 lower than that obtained from Li, Be, B and C. This is also the case for $\delta_{3}$. But the $\Delta\delta_{H}$ values estimated in PDbr-L and DR-L models remain consistent with those derived in PDbr-H and DR-H models. It seems that a same level of variation in the diffusion slope at a few hundreds GV can explain the hardening in cosmic ray spectra or ratios for both light and heavy particles. Moreover, since we assume that p and He have same diffusion slopes, the observational difference between the p and He spectra is explained by the difference between their injection indices. This can be seen in Table \ref{tab:injdiff}. For the high-energy injection parameters, the estimated $\nu_{2\text{He}}$ is lower than the $\nu_{2\text{p}}$ value in both PDbr-L and DR-L models. The $\nu_{2\text{C}}$ value in PDbr-H (or DR-H) model is estimated to be lower than the $\nu_{2\text{He}}$ value obtained in PDbr-L (or DR-L) model. It seems that above a few GV, the lighter the particle, the larger the injected spectral index.

\begin{table*}[!htbp]
\begin{spacing}{1.2}
\begin{center}
\small
\begin{tabular}{l | c | c}
\hline \hline
& PDbr-L v.s. PDbr-H &DR-L v.s. DR-H  \\
\hline
$\nu_{2\text{p}}-\nu_{2\text{He}}$  & $0.062\pm^{0.012}_{0.014}$ & $0.054\pm0.011$ \\
$\nu_{2\text{He}}-\nu_{2\text{C}}$  &  $0.027\pm^{0.013}_{0.014}$ & $0.012\pm0.014$ \\
\hline
$\nu_{1\text{p}}-\nu_{1\text{He}}$  & $0.02\pm^{0.05}_{0.06}$ & $0.289\pm0.025$\\
$\nu_{1\text{He}}-\nu_{1\text{C}}$  & $1.10\pm^{0.12}_{0.13}$ & $0.07\pm0.05$ \\
\hline
\end{tabular}
\end{center}
\end{spacing}
\caption[The injection differences]{\footnotesize The differences of injection index between various primary cosmic ray species.}\label{tab:injdiff}
\end{table*}

At low energies, a diffusion slope variation $\Delta\delta_{L}=0.64\pm^{0.05}_{0.06}$ or a \textit{Alfv$\acute{e}$n velocity} $v_{A}=12.6\pm^{1.0}_{1.1}$~km\,s$^{-1}$ is necessary to reconcile all the light nuclei data. However, while PDbr-L model agrees well with the $\bar{\text{p}}$/p data, DR-L model cannot fit the $\bar{\text{p}}$/p data below 1~GeV. This is one reason for the larger $\chi^{2}$ obtained by DR-L compared with PDbr-L model. Additionally, we calculate the predictions on antiproton flux for PDbr-L and DR-L models, as shown in Fig~\ref{fig:Z2}. It is found that both models can generally reproduce the PAMELA and AMS02 antiproton fluxes.   Nevertheless, comparing with the heavy nuclei, the low-mass data infer a smaller change in diffusion slope or a weaker reacceleration process. The low-energy injection properties are model-dependent. As shown in Table~\ref{tab:injdiff}, we find $\nu_{1\text{p}}\approx\nu_{1\text{He}}>\nu_{1\text{C}}$ for the PDbr configuration, but $\nu_{1\text{p}}>\nu_{1\text{He}}\approx\nu_{1\text{C}}$ for the DR configuration. Both results are not easy to be explained with current knowledge on acceleration mechanisms. Furthermore, as seen from Fig~\ref{fig:Z2}, both PDbr-L and DR-L models display disagreements with the PAMELA $^{3}$He/$^{4}$He ratio below 300~MeV/n and the PAMELA helium data below 400~MeV/n. This maybe because the force-field approximation is based on the zero streaming hypothesis which is only valid above 400 MeV/n~\cite{Gleeson1968}. Additionally, it is worth noting that both the light and heavy elements yield compatible values of $\phi$. It strengthens the robustness of the solar effect description for PAMELA data above 400~MeV/n.

\subsection{Crosscheck analysis}
\begin{figure}[!htbp]
\centering
\includegraphics[width=8.5cm]{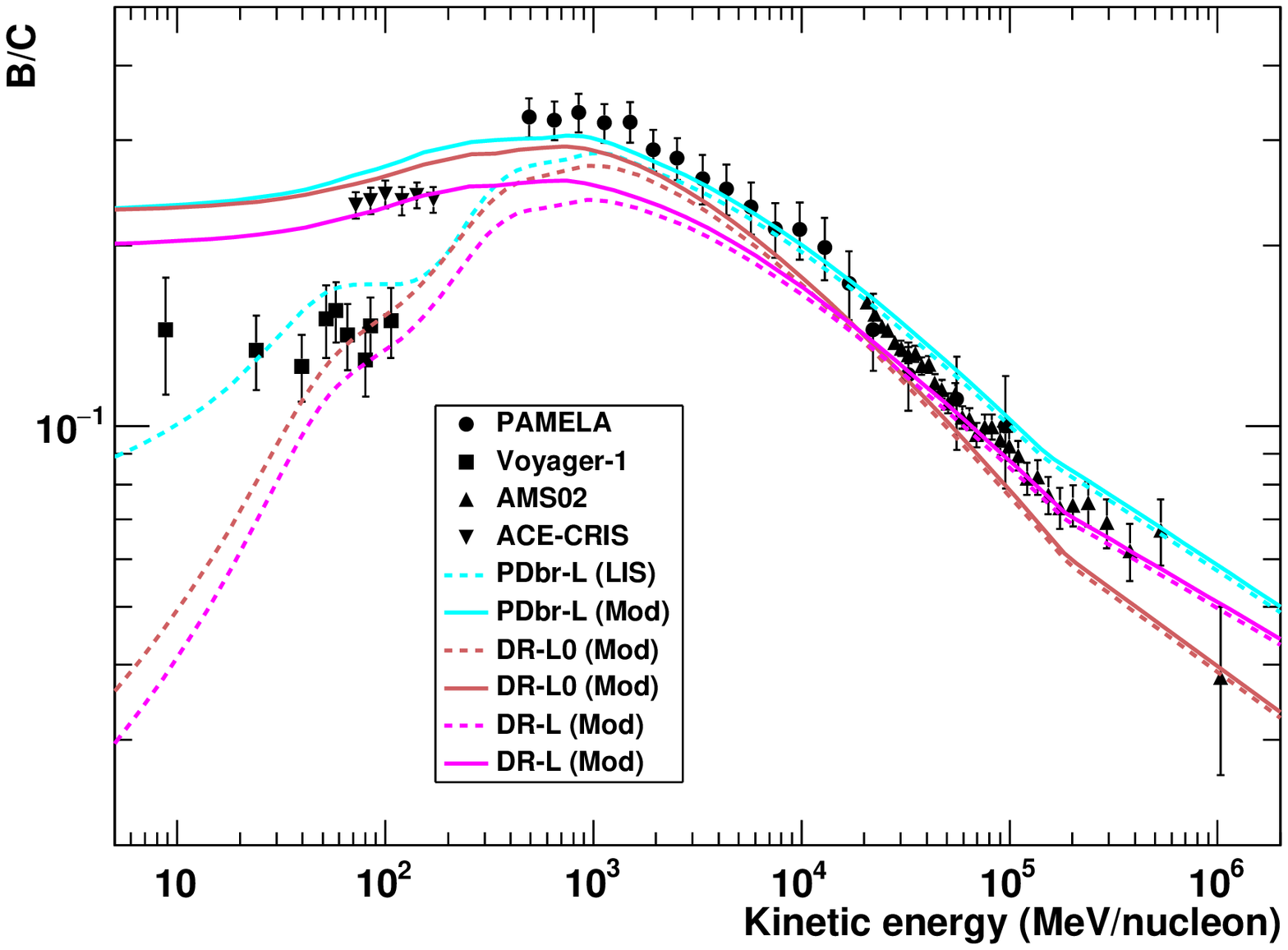}
\includegraphics[width=8.5cm]{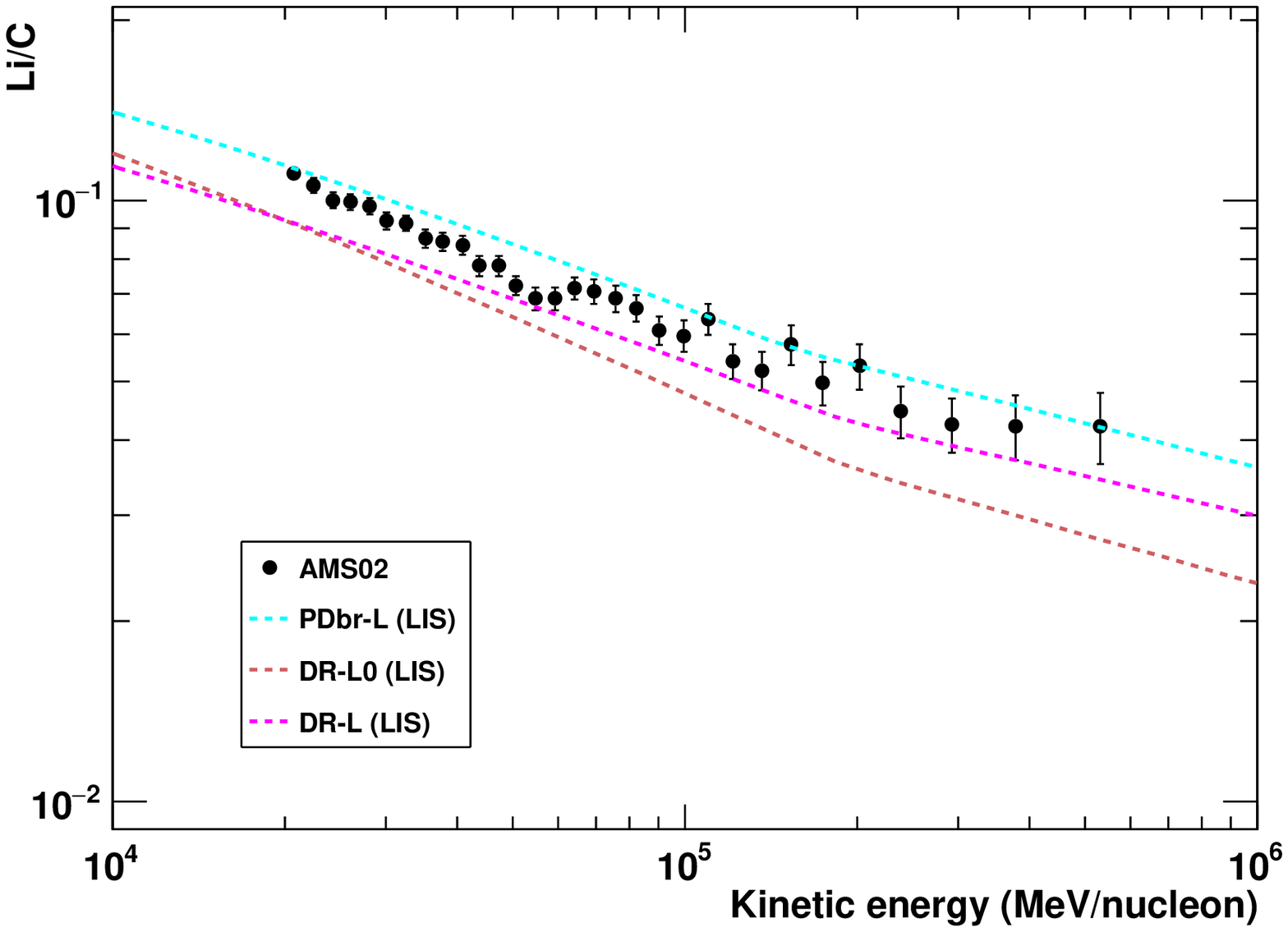}
\includegraphics[width=8.5cm]{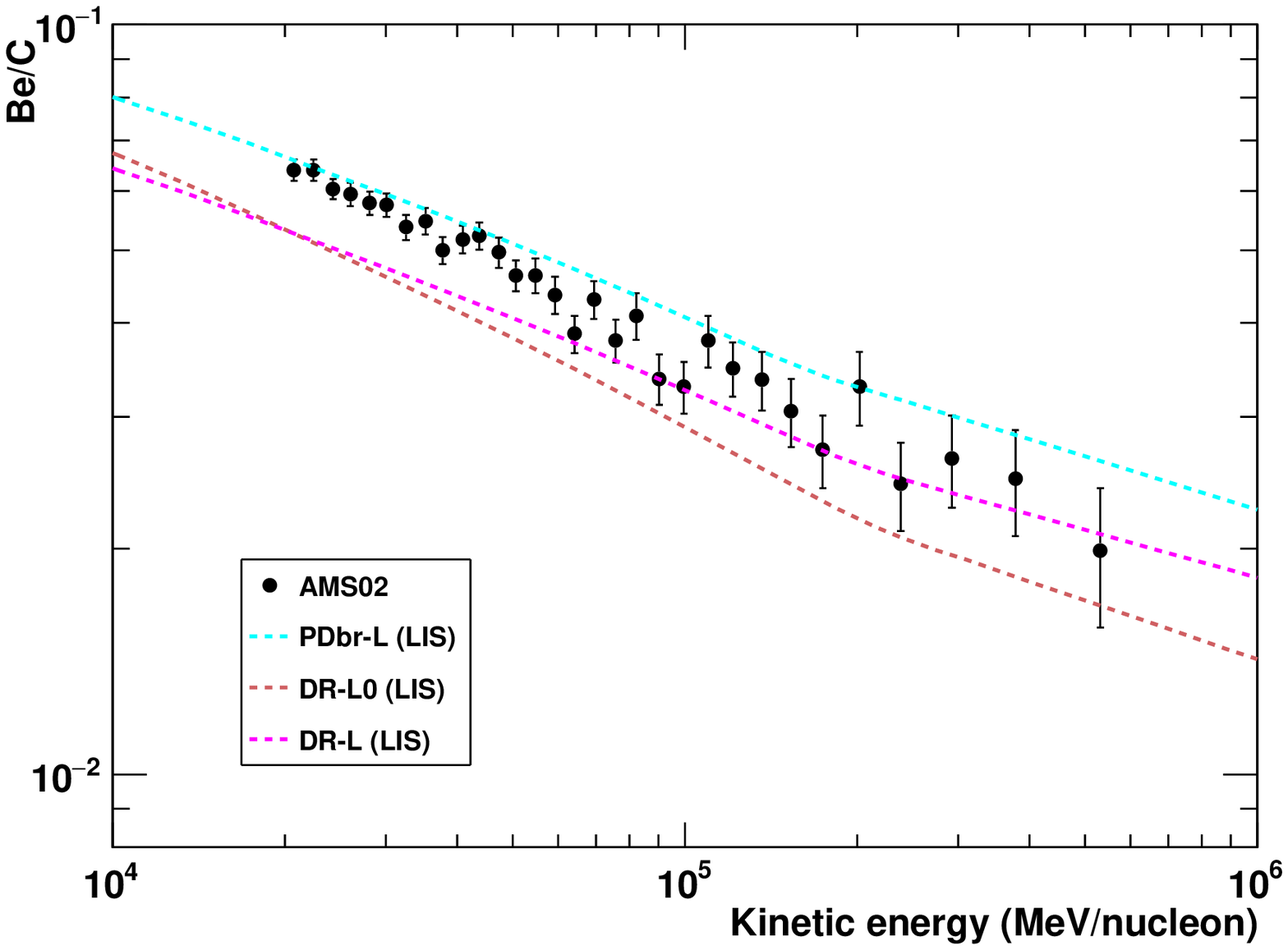}
\caption[The heavy S/P ratios based on PDbr-L and DR-L models]{\footnotesize The B/C, Li/C and Be/C ratios for the best-fit parameters of PDbr-L, DR-L0 and DR-L models as listed in Table~\ref{tab:Z2results}. The calculation from PDbr-L0 model cannot be distinguished with that from PDbr-L and is not shown here. The solid (dashed) lines represent the interstellar (modulated) spectra and ratio. Data points are the measurements from PAMELA, AMS02, ACE-CRIS and Voyager-1.}\label{fig:BCLight}
\end{figure}

In order to further understand the differences between the light and heavy nuclei, we theoretically calculate the B/C, Li/C, Be/C ratios based on the propagation and solar modulation parameters estimated in PDbr-L(0)and DR-L(0) models. The results are presented in Fig \ref{fig:BCLight}. As illustrated, no model can gives satisfactory prediction for the B/C ratio. At energies larger than 20~GeV/n, while DR-L0 model expects slightly lower B/C ratio than the AMS02 data, PDbr-L model predict higher ratio than the AMS02 observations. Compared with DR-L0 and PDbr-L models, DR-L model is closer to the AMS02 B/C data , but deviates further with the data at a few~GeV/n. At MeV range, only DR-L can fit the ACE-CRIS data. For each model, the prediction features on high-energy Li/C and Be/C ratios are similar as that on the B/C ratio. Nevertheless, none of them can explain the B/C bump around 1~GeV/n. This indicates that compared with the light nuclei, a much larger diffusion slope variation $\Delta\delta_{L}$ or a stronger $v_{A}$ is preferred to interpret the B/C peak.

\begin{figure}[!h]
\centering
\includegraphics[width=8.5cm]{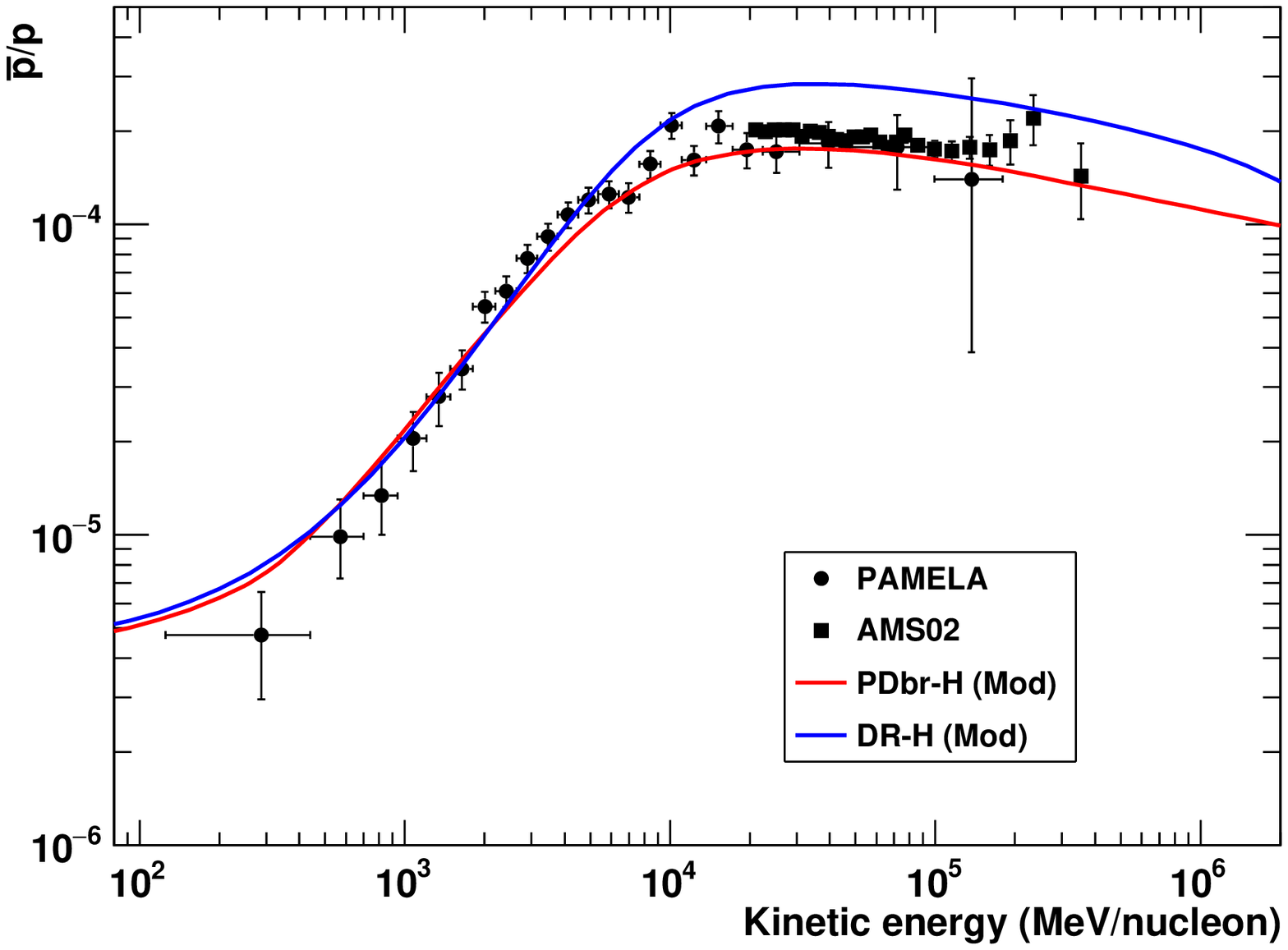}
\includegraphics[width=8.5cm]{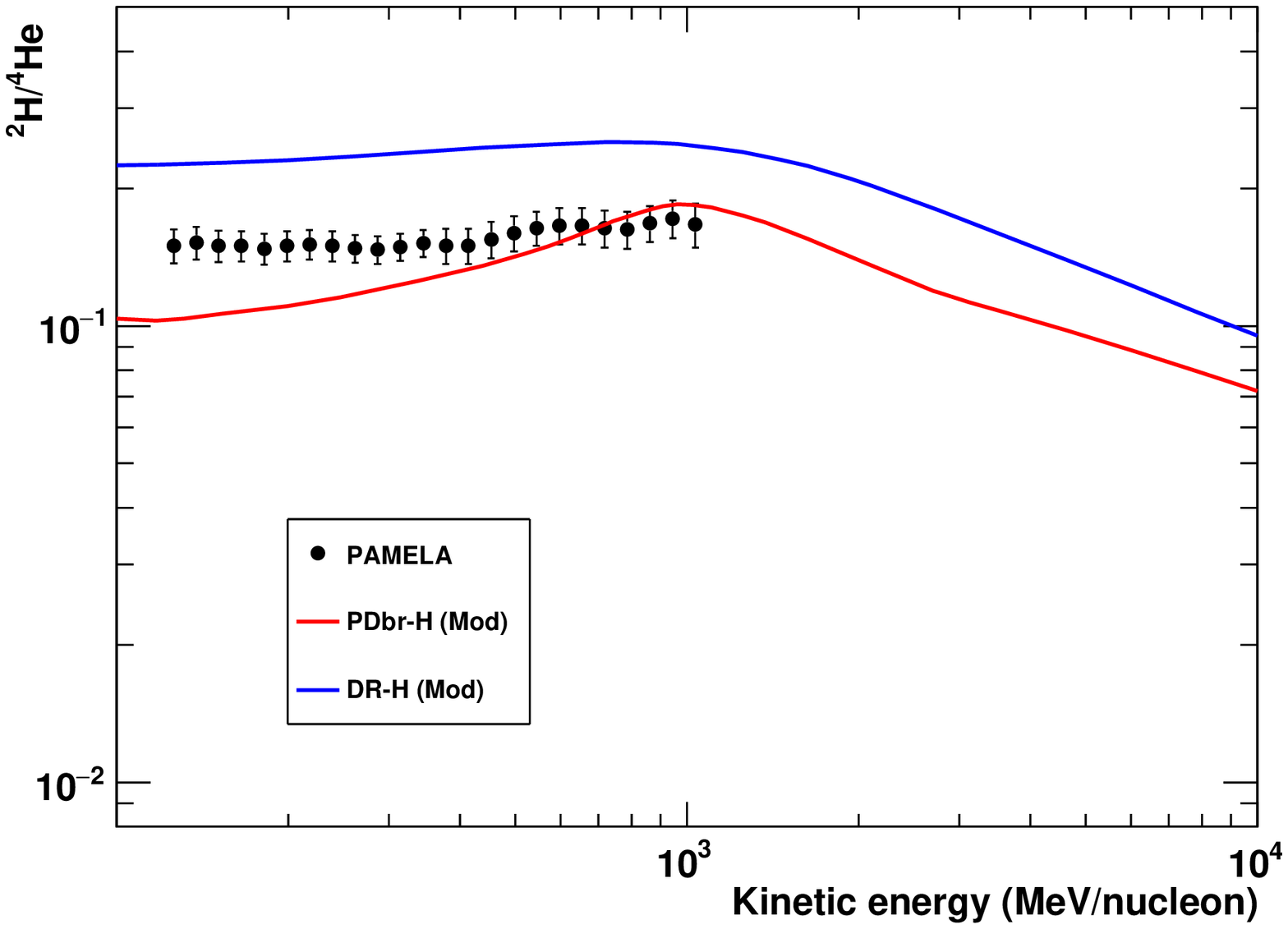}
\includegraphics[width=8.5cm]{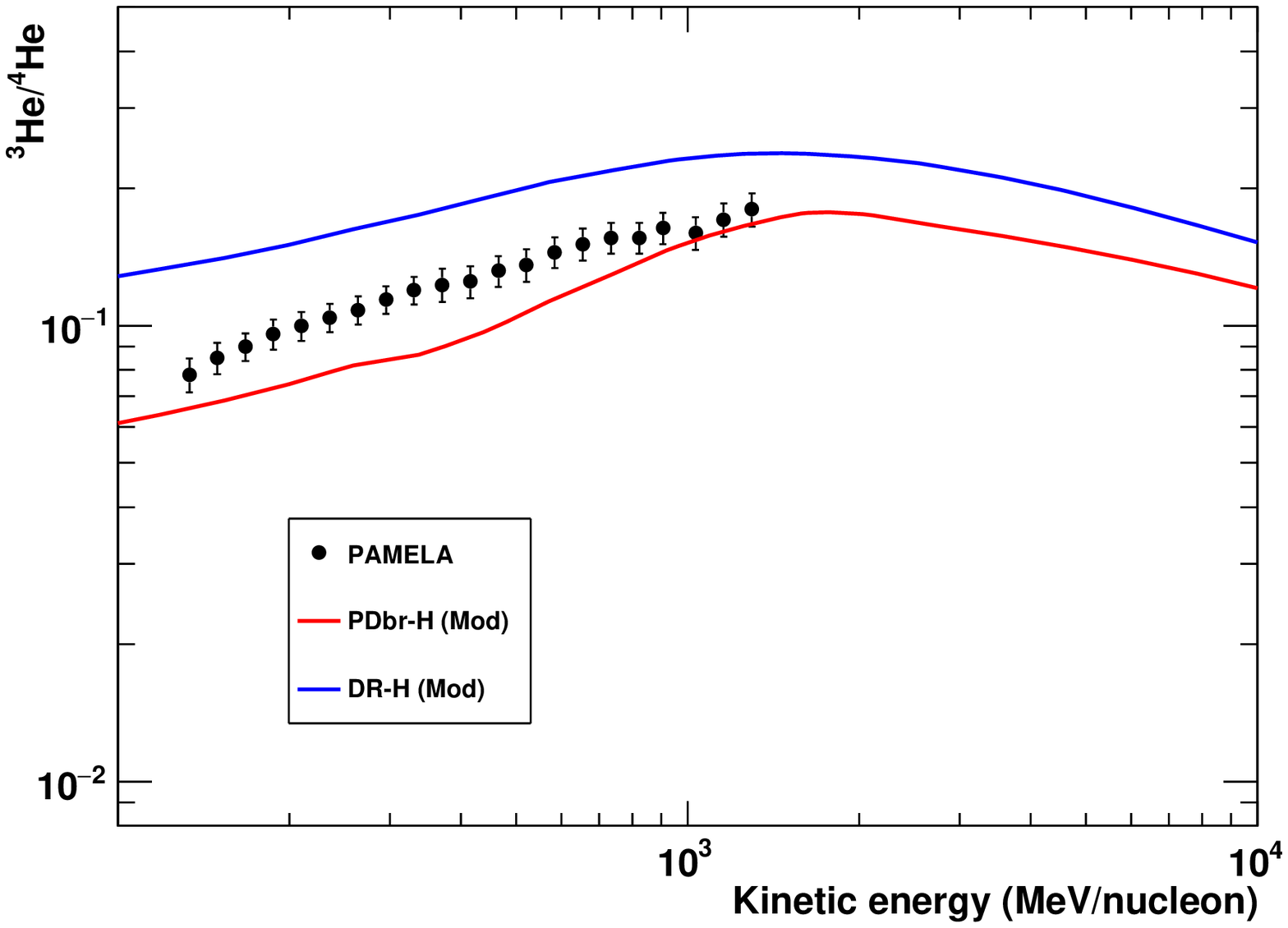}
\caption[The low-mass S/P ratios based on PDbr-H and DR-H models]{\footnotesize The $\bar{\text{p}}$/p, $^{2}$H/$^{4}$He and $^{3}$He/$^{4}$He ratios for the best-fit parameters of PDbr-H and DR-H models as listed in Table~\ref{tab:BCresults}. Data points are the measurements from PAMELA, AMS02 and Voyager-1.}\label{fig:pbarpHeisoHeavy}
\end{figure}

Subsequently, we calculate the $\bar{\text{p}}$/p, $^2$H/$^4$He and $^3$He/$^4$He ratios based on the best-fit parameters estimated in PDbr-H and DR-H models. As presented in Fig~\ref{fig:pbarpHeisoHeavy}, it seems PDbr-H model can fit $\bar{\text{p}}$/p ratio almost over all the energy range, but cannot fit the $^2$H/$^4$He and $^3$He/$^4$He data. This demonstrates that PDbr-H model can accommodate both the light and heavy nuclei above 1~GeV/n. It also indicate that we may be able to use the parameters derived from the heavy nuclei to predict antiprotons above 1~GeV/n. A clear distinction appears between DR-H model and the $\bar{\text{p}}$/p data above 10~GeV. Furthermore, both PDbr-H and DR-H models do not agree with the PAMELA $^2$H/$^4$He and $^3$He/$^4$He ratios, which further implies the possible incompatibilities between the models and the low-energy light elements. Some studies attribute the discrepancies to uncertainties in Solar modulation and (or) the antiproton cross-sections \cite{Cui2018, Cholis2019}, and correlations in data systematic errors \cite{Derome2019, Boudaud2020, Weinrich2020, Heisig2020}. These impacts will be further discussed in Section \ref{sec:discussion}.

\subsection{Comparison and Discussion}\label{sec:discussion}

Compared with our previous work~\cite{Wu2019}, in which we fitted only the PAMELA $^2$H/$^4$He, $^3$He/$^4$He, p, He data and the Voyager-1 p, He interstellar spectra, this updated study fits a more complete low-mass data by adding the $\bar{\text{p}}$/p data and the high-energy AMS02 p and He data, and meanwhile accomplish a separate analysis on the (Li, Be, B)/C and C data. When we considered different configurations, the high-energy diffusion slope $\delta_{2}$ obtained in~\cite{Wu2019} varied significantly (from 0.2 to 0.8). But in current work, we achieve a much more accurate estimation on $\delta_{2}$ between 0.38 and 0.48. This is because that the $^2$H/$^4$He and $^3$He/$^4$He ratios used in that pervious paper were only from MeV to GeV range and could not put strong constraints on the high energy propagation behavior. Moveover, a different choice on solar modulation model may affect our evaluations on $\delta_{2}$. In our previous work, we adopted a rigidity and charge-sign dependent solar modulation (CM) model~\cite{Cholis2016}, however, we find that the CM model cannot accommodate ACE-CRIS B/C and C data at MeV range and is not incorporated in current paper.

It is interesting to compare our results with some studies. The analysis in~\cite{Boschini2020}, in which they interfaced Galprop and HelMod~\cite{Boschini2018, Boschini2019} to implemented analysis on the heavy nuclei, found $\delta_{2}=0.415\pm0.025$. This value reconciles with our estimations on $\delta_{2}\sim\left(0.43, 0.48\right)$ based on Li, Be, B and C. It is worth noted that in our work a scaling factor of Li around 1.2 is required to reproduce the Li production. If we choose $S_{\text{Li}}=1$, an excess on Li will be observed. This is consistent with their results, which might be associate with the primary component of Li. A similar work was performed in~\cite{Korsmeier2021}. The authors combined Galprop and the force-field approximation to study Li, Be, B, C, N and O. They found values of $\delta_{2}=0.414\pm^{0.013}_{0.005}$, $\delta_{3}=0.271\pm^{0.026}_{0.007}$ and $v_{A}=24.04\pm^{0.91}_{2.90}$~km\,s$^{-1}$ in the DR framework, which are quite close with the values $\delta_{2}=0.446\pm0.014$, $\delta_{3}=0.29\pm^{0.03}_{0.04}$ and $v_{A}=17.2\pm^{1.8}_{1.9}$~km\,s$^{-1}$ in our DR-H model. For the PDbr framework, they found $\delta_{2}=0.48\pm^{0.01}_{0.03}$ and $\delta_{3}=0.33\pm^{0.02}_{0.03}$, which are consistent with our results $\delta_{2}=0.472\pm^{0.013}_{0.012}$ and $\delta_{3}=0.310\pm^{0.026}_{0.027}$ given in PDbr-H model. But they yielded a smaller value of $\Delta\delta_{L}=1.2\pm^{0.1}_{0.3}$ than ours. The same issue was suffered in other studies~\cite{Genolini2019, Vittino2019}, in which they also presented a weaker shift on the diffusion slope at a few GV. Several reasons could be responsible for these differences. First, the lack of low-energy ACE-CRIS data in their analysis could affect the estimation on the level of low-energy diffusion variation. Second, the use of the force-field approximation in their work might give inaccurate calculation on the AMS02 data gathered in a polarity reversal stage of HMF, as demonstrated in~\cite{Corti2019}. Third, they included nuisance cross-section parameters for all the species from Li to N, which was motivated to reduce the cross-section uncertainties~\cite{Genolini2018, Weinrich2020}, but might also eliminate the features that the data exhibit. Another work in \cite{Yuan2020} also used the heavy elements to constrain propagation models. We both adopt a scaling factor for Li. Their work found a $S_{\text{Li}}\sim 1.2$ which is compatible with ours. In their diffusion-convection model with a negligible convection effect, they yielded consistent $\delta_{2}$ with our result in PDbr-H model. But for the DR framework, their $\delta_{2}=0.362\pm0.004$ and $v_{A}=33.76\pm0.67$~km\,s$^{-1}$ differ with our results. These disagreements might be ascribed to our different treatments on the injection parameters and the diffusion slope at high energies. In our work, a high-energy break in the diffusion slope is assumed to account for the hardening of cosmic ray spectra around hundreds of GV. But in their work, they used a non-parametrized method~\cite{Ghelfi2016} to determine the interstellar primary fluxes, which might attribute the hardening of the energy spectrum to the hardening of the injection spectrum. Consequently, this could result in their stronger reacceleration effect and lower diffusion slope than ours. Note that their larger $v_{A}$ could also explain the stronger hardening in the secondary spectra than in the primary ones.

Our results derived from light particles are compared with the results given in \cite{Cuoco2019}. In that paper, they used Galprop to study p, he and $\bar{\text{p}}$/p. A main difference between our work is that we use the low-energy PAMELA light nuclei data collected during the solar minimum period, while they utilized only data with rigidities larger than 5~GV. The $\delta_{2}$ and $v_{A}$ values determined in our DR-L0 model are $0.462\pm0.010$ and $15.8\pm^{0.7}_{0.8}$~km\,s$^{-1}$ respectively. While our $v_{A}$ estimation is consistent with theirs, our $\delta_{2}$ value is slightly higher than their $\delta_{2}=0.42\pm^{0.02}_{0.01}$. This difference might be caused by the fact they only adopt data $>5$~GV. And their assumption made on the high energy diffusion slope, i.e. $\Delta\delta_{H}=-0.12$, could also influence the determination on $\delta_{2}$.

Some studies \cite{Reinert2018, Evoli2019, Cholis2019, Mertsch2020} found that the heavy and light particles can be explained with identical propagation mechanisms. But in their work, they either didn't employ the low-energy B/C, $^2$H and $^3$He data, or only considered the high energy particles. This fact is reconcilable with our results, since our PDbr-H models can generally reproduce all the data above 1~GeV/n. A study in \cite{Cui2018} found that in the DR framework the propagation parameters estimated from the heavy nuclei (including the ACE-CRIS B/C data) can fit the $\bar{\text{p}}$/p ratio. This seems disagree with our results, since our DR-H models based on the heavy elements face difficulties in reproducing the $\bar{\text{p}}$/p ratio. The reason to explain this difference could be the use of a rescaling factor on antiproton production cross-sections in their papers, which might diminish the discrepancies between the $\bar{\text{p}}$ and B/C data. But we do not expect this factor to be responsible for the contradiction between DR-H model and the $^2$H, $^3$He data.

Another recent work combined the AMS02 $^3$He and PAMELA $^3$He/$^4$He data with heavy particles in the analysis~\cite{Weinrich2020}. For different combination of data sets, they provided $\delta_{2}$ around 0.51 under the PDbr configuration and 0.47 under the DR configuration. Generally, they obtained slightly higher values on $\delta_{2}$ than ours. The difference might caused by several reasons. First, they used the simplified analytical approach for propagation, but we use the fully numerical Galprop code. Second, they used the force-field approximation to describe AMS02 data, but we use PAMELA data instead of the AMS02 data below 20~GeV/n to diminish the uncertainties in Solar modulation. Besides, they mainly focused on analysis on secondaries, while we include the primaries in the fits. Finally, they introduced a correlation matrix to handle experimental data errors and took into account uncertainties in nuclear cross-sections not only for Li. All these factors could influence the best-fit parameters. Though they reported all the data they employed can be reproduced by their studied models, they stated this result is strongly impacted by the treatment on systematic correlations. Nevertheless, a consistent conclusion was found with ours, that is, with an inclusion of $^3$He data in the analysis, the best-fit $\Delta\delta_{L}$ is lower than the result obtained only from (Li, Be, B)/C. And an inclusion of the ACE-CRIS B/C data may leads larger $\Delta\delta_{L}$ than only using the AMS02 B/C data.

\section{Conclusion}

In this work, we use the $Z>2$ nuclei and the Z$\leq$2 elements separately to substantially study two cosmic ray acceleration and propagation models. One is the plain diffusion model with a low-energy break in the diffusion coefficient, the other is the diffusion-reacceleration model. Our results rely on four different combinations of data sets. For both heavy and light particles, different data-set combinations achieve consistent evaluations on the cosmic ray acceleration and propagation parameters under the plain diffusion framework, but yield significant discrepant results under the diffusion-reacceleration framework. Nonetheless, the heavy elements put constraints on $\delta_{2}\sim\left(0.42, 0.48\right)$ and $\delta_{3}\sim\left(0.22, 0.34\right)$, while the light species give $\delta_{2}\sim\left(0.38, 0.47\right)$ and $\delta_{3}\sim\left(0.21, 0.30\right)$. It seems that the $\bar{\text{p}}$/p ratio put a looser restriction on $\delta_{2}$ but a stronger constraint on $\delta_{3}$ than the (Li, Be, B)/C ratios. But their results show overlaps on $\delta_{2}$ and $\delta_{3}$. Moreover, the $\Delta\delta_{H}$ values are determined around $-0.18\sim-0.15$ for both the light and heavy particles. All these facts indicate that the light and heavy nuclei can yield compatible results at medium to high energies. All the particles above 1~GeV/n can be accommodated in same models. The re-normalization factor of Li with values other than 1 may be related either with the possible primary component, or with an improper normalization on cross-sections of Li.

At low energies, the ACE-CRIS B/C data and the PAMELA $^2$H/$^4$He, $^3$He/$^4$He are all sensitive to the low-energy parameters. Compared with the $^2$H/$^4$He, $^3$He/$^4$He data, we find that the ACE-CRIS B/C ratio requires a more dramatic change on the diffusion slope at a few GV, or a stronger reacceleration. Such discrepancies infer that the light nuclei may suffer divergent low-energy transport behaviors with the heavy particles, which may challenge our traditional understanding on cosmic rays. However, in this work we fix the halo size $z_{h}$ to 4~kpc. This may lead underestimations on the errors for other free parameters and may affect our conclusion. Precise measurements on radioactive species such as $^{10}$Be are awaited to put on a stringent constraint on $z_{h}$. As discussed in Section~\ref{sec:discussion}, uncertainties in cross-sections and correlations in data systematical errors, may also impact the results. More accurate estimations on the cross-sections and on the systematical correlations can help us better clarify the consistency/inconsistency between the heavy and light cosmic rays. Moreover, the convection process is not considered in this paper, which may influence our results, especially on the low-energy behaviors for both light and heavy nuclei.

Furthermore, considering the limited precisions of the PAMELA data below 20~GeV/n, if the solar modulation during the reversal period of HMF can be treated reliably, the accurate AMS02 low-energy data may provide more helpful insights on the properties of cosmic rays. Particularly, while PAMELA measured $^2$H/$^4$He and $^3$He/$^4$He ratios only from MeV to GeV range, AMS02 have provided the $^3$He/$^4$He data and will publish the $^2$H/$^4$He ratio extending to 10~GeV/n which may allow us to extract more rigorous and accurate constraints on cosmic ray propagation. Further efforts on examining a robust solar modulation model for AMS02 data are needed to strengthen our understanding on the cosmic ray acceleration and propagation mechanisms.

\section{Acknowledgements}

We thank Michael Korsmeier, Su-jie Lin and Qiang Yuan for very helpful discussions. This work is supported by the Joint Funds of the National Natural Science Foundation of China (Grant No. U1738130). The use of the high-performance computing platform of China University of Geosciences is gratefully acknowledged.

\end{document}